\documentclass[twocolumn]{aastex63}

\newcommand{\specz}{z_\mathrm{spec}}
\newcommand{\photz}{z_\mathrm{phot}}

\newcommand{\mstar}{M_{\star}}
\newcommand{\msun}{M_{\odot}}
\newcommand{\mbh}{M_\mathrm{BH}}

\newcommand{\mdotbh}{\dot{M}_\mathrm{BH}}

\newcommand{\lbol}{L_\mathrm{AGN,bol}}
\newcommand{\ledd}{L_\mathrm{Edd}}

\newcommand{\liragn}{L_\mathrm{AGN,IR}}
\newcommand{\liragnunit}{\liragn/\mathrm{erg}~\mathrm{s}^{-1}}

\newcommand{\etar}{\eta_\mathrm{rad}}

\newcommand{\lfirst}{L_\mathrm{1.4GHz}}

\newcommand{\ljet}{L_\mathrm{jet}}
\newcommand{\ljetunit}{L_\mathrm{jet}/\mathrm{erg}~\mathrm{s}^{-1}}
\newcommand{\etaj}{\eta_\mathrm{jet}}

\newcommand{\lambdaedd}{\lambda_\mathrm{Edd}}

\newcommand{\R}{\mathcal{R_{\rm rest}}}
\newcommand{\Rint}{\mathcal{R_\mathrm{int}}}
\newcommand{\whz}{\mathrm{W}~\mathrm{Hz}^{-1}}
\newcommand{\sBHARunit}{\mathrm{erg}~\mathrm{s}^{-1}~M_\odot^{-1}}
\newcommand{\SFR}{\mathrm{SFR}}
\newcommand{\sSFR}{\mathrm{sSFR}}
\newcommand{\qir}{q_\mathrm{IR}}
\newcommand{\fbstar}{f_\mathrm{b,\star}}

\newcommand{\NRGs}{NRGs}
\newcommand{\ERGs}{ERGs}

\newcommand{\Nergs}{39} 
\newcommand{\Nwergstot}{1026} 
\newcommand{\Nwergsother}{987} 

\usepackage{natbib}
\usepackage{amsmath}
\usepackage{multirow} 
\usepackage{ifthen} 

\received{May 3, 2020}
\revised{July 19, 2021}
\accepted{August 5, 2021}
\submitjournal{ApJ}

\shorttitle{Extremely radio loud galaxies in WERGS}
\shortauthors{Ichikawa et al.}

\begin{document}



\title{
A Wide and Deep Exploration of Radio Galaxies with Subaru HSC (WERGS). IV.
Rapidly Growing (Super-)Massive Black Holes in Extremely Radio-Loud Galaxies
}

\correspondingauthor{Kohei Ichikawa}
\email{k.ichikawa@astr.tohoku.ac.jp}

\author[0000-0002-4377-903X]{Kohei Ichikawa}
\affil{Frontier Research Institute for Interdisciplinary Sciences, Tohoku University, Sendai 980-8578, Japan}
\affil{
Astronomical Institute, Tohoku University, Aramaki, Aoba-ku, Sendai, Miyagi 980-8578, Japan
}
\affil{Max-Planck-Institut f{\"u}r extraterrestrische Physik (MPE), Giessenbachstrasse 1, D-85748 Garching bei M{\"u}unchen, Germany
}

\author[0000-0002-4999-9965]{Takuji Yamashita}
\affil{
National Astronomical Observatory of Japan, 2-21-1 Osawa, Mitaka, Tokyo 181-8588, Japan
}

\author[0000-0002-3531-7863]{Yoshiki Toba}
\affil{
Department of Astronomy, Kyoto University, Kitashirakawa-Oiwake-cho, Sakyo-ku, Kyoto 606-8502, Japan
}
\affil{
Academia Sinica Institute of Astronomy and Astrophysics, 11F of Astronomy-Mathematics Building, AS/NTU, No.1, Section 4, Roosevelt Road, Taipei 10617, Taiwan
}
\affil{
Research Center for Space and Cosmic Evolution, Ehime University, 2-5 
Bunkyo-cho, Matsuyama, Ehime 790-8577, Japan
}

\author[0000-0002-7402-5441]{Tohru Nagao}
\affil{
Research Center for Space and Cosmic Evolution, Ehime University, 2-5 
Bunkyo-cho, Matsuyama, Ehime 790-8577, Japan
}

\author[0000-0001-9840-4959]{Kohei Inayoshi}
\affil{
Kavli Institute for Astronomy and Astrophysics, Peking University, Beijing 100871, China
}

\author{Maria Charisi}
\affil{
TAPIR, California Institute of Technology, 1200 E. California Blvd., Pasadena, CA 91125, USA}

\author[0000-0001-7759-6410]{Wanqiu He}
\affil{
Astronomical Institute, Tohoku University, Aramaki, Aoba-ku, Sendai, Miyagi 980-8578, Japan
}

\author[0000-0002-5104-6434]{Alexander Y. Wagner}
\affil{
Center for Computational Sciences, University of Tsukuba, 1-1-1 Tennodai, Tsukuba, Ibaraki 305-8577, Japan
}

\author[0000-0002-2651-1701]{Masayuki Akiyama}
\affil{
Astronomical Institute, Tohoku University, Aramaki, Aoba-ku, Sendai, Miyagi 980-8578, Japan
}

\author{Bovornpratch Vijarnwannaluk}
\affil{
Astronomical Institute, Tohoku University, Aramaki, Aoba-ku, Sendai, Miyagi 980-8578, Japan
}

\author[0000-0003-2682-473X]{Xiaoyang Chen}
\affil{
Astronomical Institute, Tohoku University, Aramaki, Aoba-ku, Sendai, Miyagi 980-8578, Japan
}

\author[0000-0002-1732-6387]{Masaru Kajisawa}
\affil{
Research Center for Space and Cosmic Evolution, Ehime University, 2-5 
Bunkyo-cho, Matsuyama, Ehime 790-8577, Japan
}

\author[0000-0002-6808-2052]{Taiki Kawamuro}
\affil{
National Astronomical Observatory of Japan, 2-21-1 Osawa, Mitaka, Tokyo 181-8588, Japan
}

\author[0000-0003-1700-5740]{Chien-Hsiu Lee}
\affil{NSF's National Optical-Infrared Astronomy Research Laboratory, Tucson, AZ 85719, USA}

\author{Yoshiki Matsuoka}
\affil{
Research Center for Space and Cosmic Evolution, Ehime University, 2-5 
Bunkyo-cho, Matsuyama, Ehime 790-8577, Japan
}

\author{Malte Schramm}
\affil{
National Astronomical Observatory of Japan, 2-21-1 Osawa, Mitaka, Tokyo 181-8588, Japan
}

\author[0000-0002-2536-1633]{Hyewon Suh}
\affil{
Subaru Telescope, National Astronomical Observatory of Japan (NAOJ), 650 North A'ohoku Place, Hilo, HI 96720, USA
}
\affil{
Gemini Observatory/NSF's NOIRLab, 670 N. A'ohoku Place, Hilo, Hawaii, 96720, USA
}

\author[0000-0002-5011-5178]{Masayuki Tanaka}
\affil{
National Astronomical Observatory of Japan, 2-21-1 Osawa, Mitaka, Tokyo 181-8588, Japan
}

\author{Hisakazu Uchiyama}
\affil{
National Astronomical Observatory of Japan, 2-21-1 Osawa, Mitaka, Tokyo 181-8588, Japan
}

\author[0000-0001-7821-6715]{Yoshihiro Ueda}
\affil{
Department of Astronomy, Kyoto University, Kitashirakawa-Oiwake-cho, Sakyo-ku, Kyoto 606-8502, Japan
}

\author{Janek Pflugradt}
\affil{
Astronomical Institute, Tohoku University, Aramaki, Aoba-ku, Sendai, Miyagi 980-8578, Japan
}

\author{Hikaru Fukuchi}
\affil{
Astronomical Institute, Tohoku University, Aramaki, Aoba-ku, Sendai, Miyagi 980-8578, Japan
}




\begin{abstract}

We present the optical and infrared properties of
\Nergs\ extremely radio-loud galaxies discovered by 
cross-matching the Subaru/Hyper Suprime-Cam (HSC) 
deep optical imaging survey and VLA/FIRST 1.4~GHz radio survey.
The recent Subaru/HSC strategic survey revealed optically-faint radio galaxies (RG) down to $g_\mathrm{AB} \sim26$, 
opening a new parameter space of extremely radio-loud 
galaxies (ERGs) with radio-loudness parameter of
$\log \R = \log (f_{1.4 \mathrm{ GHz,rest}}/f_{g,\mathrm{rest}}) >4$.
Because of their optical faintness and small
number density of $\sim1~$deg$^{-2}$,
such ERGs were difficult to find in the previous wide but shallow, or deep but small area optical surveys.
ERGs show intriguing properties that are different from the conventional RGs:
(1) most ERGs reside above or on the star-forming main-sequence, and some of them might be low-mass galaxies with $\log (\mstar/\msun) < 10$. 
(2) ERGs exhibit a high specific black hole accretion rate, reaching the order of the Eddington limit.
The intrinsic radio-loudness ($\Rint$), defined by the ratio of jet power over bolometric radiation luminosity, is one order of magnitude higher than that of radio quasars. This suggests that ERGs harbor a unique type of active galactic nuclei (AGN) that show both powerful radiations and jets.
 Therefore, ERGs are prominent candidates of very rapidly growing black holes reaching Eddington-limited accretion just before the onset of intensive AGN feedback.

\end{abstract}

\keywords{galaxies: active --- 
galaxies: nuclei ---
quasars: supermassive black holes ---}



\section{Introduction}\label{sec:intro}

The formation of supermassive black holes (SMBHs) and their growth across the cosmic time are fundamental questions in modern astronomy.
In the local universe at $z\sim0$, 
the mass of SMBH ($\mbh$) and
their host properties show tight correlations
for the SMBH mass range of $6 \lesssim \log (\mbh/\msun) \lesssim 10$
with a scatter of $\sigma \sim 0.3$~dex \citep[e.g.,][]{mag98,geb00,fer00,tre02,mar03,har04,san11,kor13,mcc13}.
Such a tight correlation is considered to be established by a balance of feeding and feedback processes between the central SMBHs and the host galaxies.

Local Radio galaxies have been primary targets
for investigating the effect of SMBHs on the host galaxies because powerful radio galaxies or radio-loud active galactic nuclei (AGN) mainly reside in
massive galaxies whose star-formation is quenched,
with the presence of strong jets dispersing the interstellar medium \citep[e.g.,][]{mor05, hol08, nes17}, or producing cavities in the host galaxies \citep[e.g.,][]{raf06,mcn07,bir08,bla19}.
Those radio galaxies tend to show a low accretion rate, i.e., Eddington ratio of $\lambda_\mathrm{Edd} < 10^{-2}$, 
suggesting that the energy release is dominated by the kinetic power by jet, not by the radiation from AGN accretion disk.

However, the situation may be different at $z>1$. 
Using over $10^3$ radio AGN selected from the
Very Large Array (VLA)-COSMOS 3~GHz large project \citep{smo17a}, \cite{del18} demonstrated that
SMBH accretion in radio-bright ($\lfirst > 10^{25}~\whz$) AGN becomes more radiatively efficient ($\lambda_\mathrm{Edd}>10^{-2}$) at $z>1$.
They reside in star-forming galaxies, which contain plenty of cold gas.
This picture of radio AGN is completely different from those seen in the local universe
in the same radio luminosity range \citep[e.g.,][]{hic09}. Still, the survey volume of VLA-COSMOS surveys is small so that they may be missing a rare, but radio-bright population.
The FIRST survey is the best tool to explore such a radio bright end since it covers half the sky
with the VLA at 1.4~GHz
\citep[e.g.,][]{bec95,hel15}. However,
cross-matching the VLA/FIRST sources with the SDSS survey catalog identified optical counterparts in only 30\% of 
the radio sources \citep{ive02,bes12}.

\begin{figure}
\begin{center}
\includegraphics[width=0.48\textwidth]{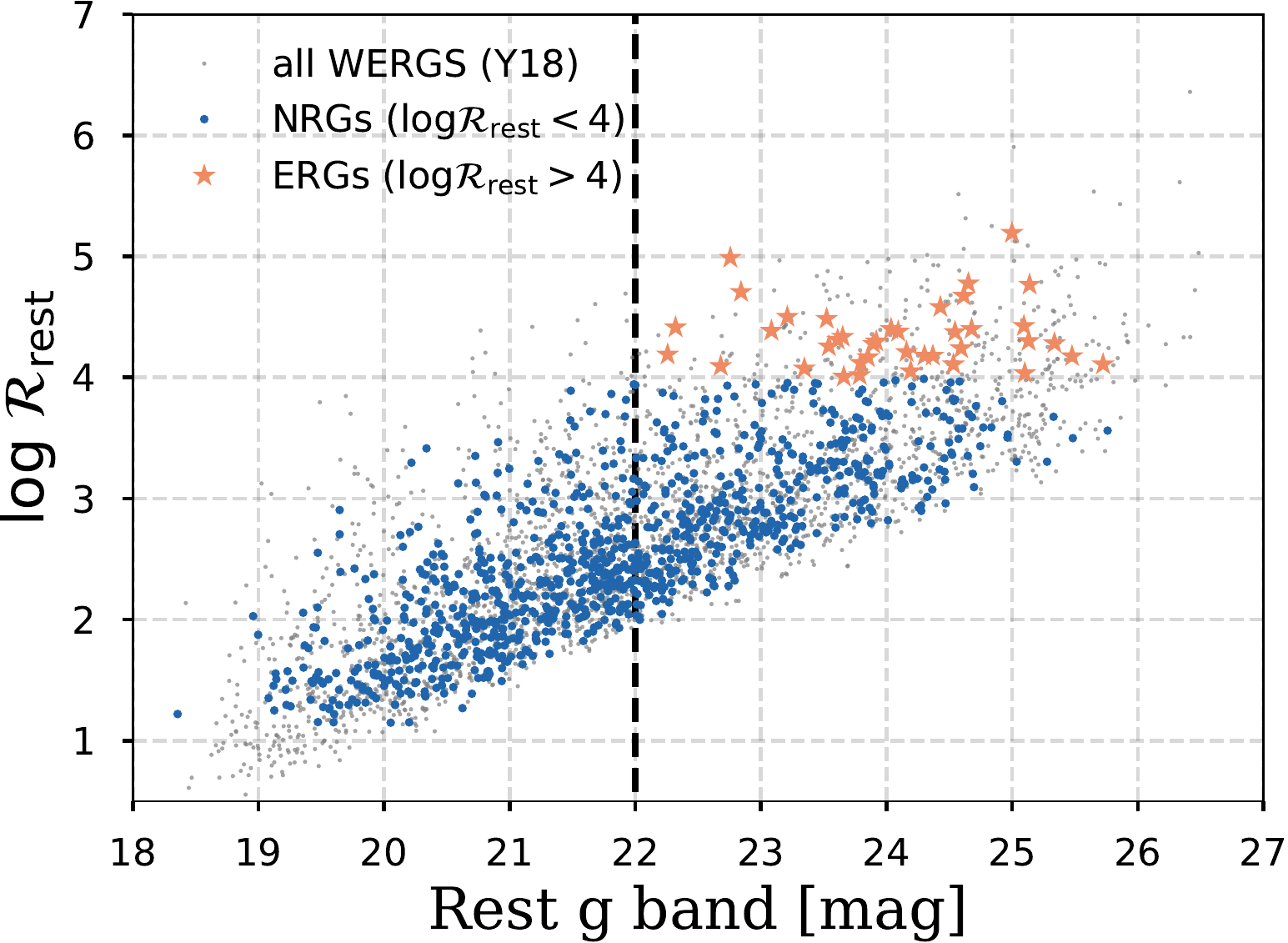}
\caption{
$\log \R$ versus rest $g$ band magnitude of our sample.
The original WERGS sample obtained by \cite{yam18} is shown with gray circles. The finally selected \Nwergsother~WERGS sample with $\log \R < 4$ used in this study (we call the sample ``\NRGs'' in the text) and \Nergs~\ERGs\ (the sample with $\log \R>4$) are shown with blue circles, and orange stars, respectively.
The vertical dashed-line indicates the magnitude limit of the SDSS survey at $g_\mathrm{AB} = 22$.
}\label{fig:R_vs_gmag}
\end{center}
\end{figure}

Recent Subaru/Hyper Suprime-Cam \citep[HSC; ][]{miy18} strategic survey shed light on such a situation. 
We have conducted a search for optically-faint radio-galaxies (RGs) using the Subaru HSC survey catalog \citep{aih18a} and the VLA/FIRST radio continuum catalog, and we have found a large number ($>3\times10^3$ sources) of RGs at $z \sim 0$--$5$ \citep{yam18,yam20}. 
The project is called \textbf{W}ide and deep \textbf{E}xploration of \textbf{R}adio \textbf{G}alaxies with \textbf{S}ubaru/HSC
\citep[\textbf{WERGS};][]{yam18}.
\cite{yam18} demonstrated that over 60\% of populations now have reliable optical counterparts thanks to deep HSC/optical imaging.
 Figure~\ref{fig:R_vs_gmag} shows
 that the WERGS sample spans a wide range
 of optical magnitude $g_\mathrm{AB}=18-26$
 and radio-loudness parameter of
$\log \R =\log ( f_{1.4 \mathrm{GHz,rest}}/f_{g~{\rm band,rest}}) =1$--$6$ \citep[e.g.,][]{ive02}.

Our particular interest is the new parameter space of optically faint RGs that are extremely radio-loud galaxies (ERGs) with $\log \R >4$.
Previously known radio quasars have
a peak of radio-loudness at $\log \R=2$--$3$, 
and rarely contain any sources with $\log \R>4$ \citep{ive02,ino17}.
Therefore, the optical emission of ERGs seems not
originate from the AGN accretion disk anymore probably due to dust obscuration of the nuclei, but it traces 
the stellar-component of the host galaxies \citep[e.g.,][]{ter03}.
Given that the optical band is very faint 
with median magnitudes of $\left< g_\mathrm{AB}\right> \sim 24.5$, ERGs might have smaller stellar mass than previously known radio sources.
Considering the very small number density of ERGs ($\sim1$ source~deg$^{-2}$), it is not surprising that deep but smaller volume surveys
 \citep[such as VLA-COSMOS][]{smo17a,smo17b} have rarely detected such ERGs, and the very wide SDSS survey \citep[e.g., SDSS-selected radio galaxies][]{bes12} could not detect
 ERGs either because of their shallow sensitivity down to only $g_\mathrm{AB} \sim 22$.

In this study, we will explore the physical properties of ERGs.
\cite{tob19a} already compiled multi-wavelength data covering the optical, IR, and radio band for the WERGS sample,
and derived physical parameters of host galaxies such as stellar mass, star-formation rate as well as the AGN luminosity estimated from the mid-IR bands.
By using this data set, we will explore the AGN and host galaxy properties of the ERGs.
Thoroughout this paper, we 
adopt the same cosmological parameters 
as \cite{yam18} and \cite{tob19a}; $H_0 = 70$~km~s$^{-1}$~Mpc$^{-1}$, $\Omega_\mathrm{M}=0.27$, and $\Omega_\Lambda=0.73$.

\begin{figure*}
\begin{center}
\includegraphics[width=0.48\textwidth]{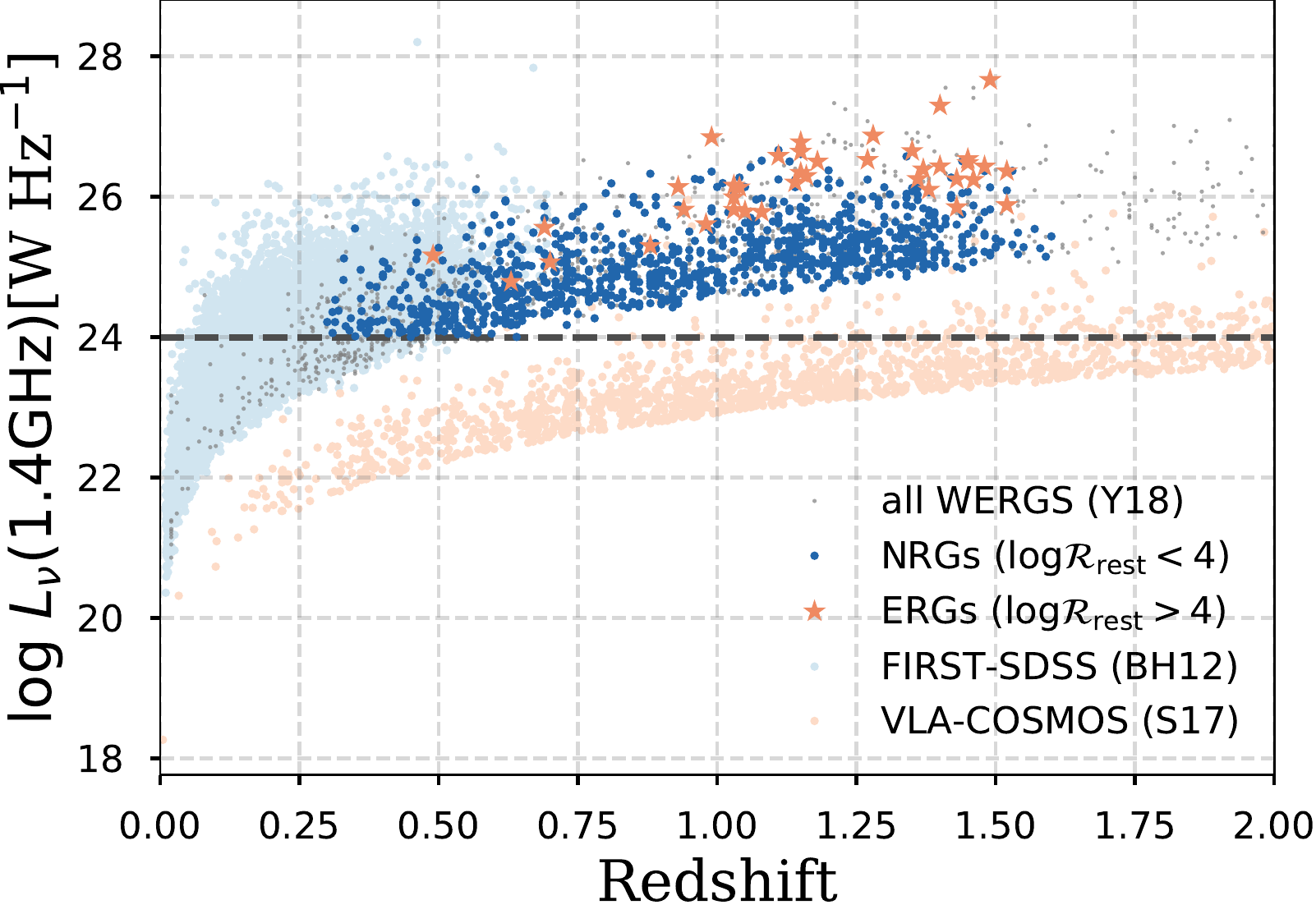}~
\includegraphics[width=0.48\textwidth]{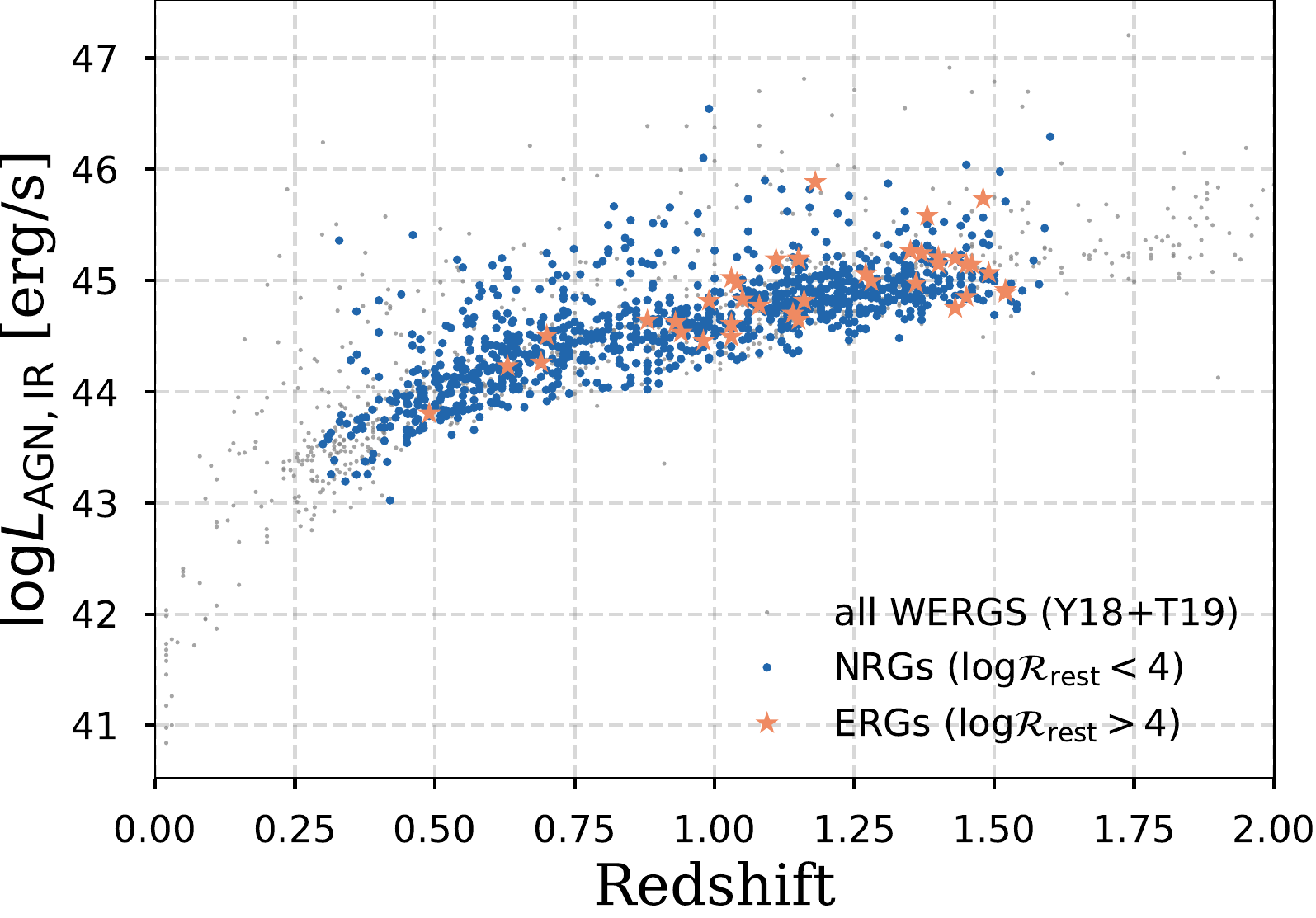}~
\caption{
Luminosity of WERGS sample as a function of 
redshift. The points are the same as in Figure~\ref{fig:R_vs_gmag}.
(left) Rest-frame
radio luminosity $\log \lfirst$ ($\whz$) as a function of redshift, where $L_\mathrm{1.4GHz}$ is obtained from VLA/FIRST \citep{hel15},
and the $k$-correction was made by \cite{yam18} based on redshift.
The horizontal dashed line is a luminosity cut we adopt to exclude galaxies with strong starbursts.
The faint cyan and orange circle points are data obtained from FIRST-SDSS \citep{bes12} and VLA-COSMOS \citep{smo17a}, respectively.
(Right) IR AGN luminosity $\liragn$ (erg~s$^{-1}$) as a function of redshift. For the entire WERGS population,
we plot the sources only when it
is included in \cite{tob19a}, in which the IR AGN luminosity was measured.
}\label{fig:L_vs_z}
\end{center}
\end{figure*}

\section{Sample Selection and Properties}\label{sec:sample}

Our initial sample is based on the WERGS sample \citep[$\sim 3600$ sources;][]{yam18}, 
which compiled the 
HSC Subaru Strategic Program (HSC-SSP)
 optical counterparts of the VLA/FIRST data.
Here, we briefly summarize the WERGS sample and the
reader should refer to \cite{yam18} for the WERGS catalog and 
\cite{tob19a} for the IR catalog of the WERGS sample.
We also describe further selection criteria imposed on the sample, which are suitable for this study. The number of WERGS sources after each sample selection cut is also summarized in Table~\ref{tab:selection}.

\subsection{WERGS Sample}\label{sec:wergs}

\subsubsection{VLA/FIRST}\label{sec:first}
The VLA/FIRST survey contains radio imaging
data at 1.4~GHz with a spatial resolution of 5.4~arcsec \citep{bec95,whi97}, 
which completely covers the footprint of the HSC-SSP Wide-layer (see the text below).
\cite{yam18} utilized the final release catalog of FIRST \citep{hel15} with the flux limit of $>1$~mJy,
and extracted 7072 FIRST sources in the HSC-SSP footprint.

\subsubsection{Subaru/HSC-SSP data}\label{sec:hsc}
The HSC-SSP is an ongoing wide and deep imaging survey covering
five broadband filters \citep[$g$-, $r$-, $i$-, $z$-, and $y$-band;][]{aih18a,bos18,fur18,kaw18,kom18,hua18},
consisting of three layers (Wide, Deep, and UltraDeep).
\cite{yam18} utilized the Wide-layer S16A data release,
which contains observations from 2014 March to 2016 January with a field coverage of 154 deg$^2$ (based on six fields;
XMM-LSS, GAMA09H, WIDE12H, GAMA15H, HECTOMAP, and VVDS), and
 forced photometry of $5\sigma$ limiting magnitude down to 
$26.8, 26.4, 26.4, 25.5,$ and $24.7$ for $g, r, i, z,$ and $y$-band, 
respectively \citep{aih18b}. 
The average seeing in the $i$-band is $0.6$ arcsec, 
and the astrometric root mean squared uncertainty is about 40~mas.
After removing spurious sources flagged by the pipeline,
\cite{yam18} cross-matched the FIRST sources with a search radius of 1~arcsec. 
\cite{yam18} also required detections with $\mathrm{S/N}>5$ in the $r$-, $i$-, and $z$-bands to qualify as an optical counterpart.
This initial sample of \cite{yam18} comprised 3579 sources.

We then applied additional cuts to this sample.
To remove  radio galaxies from the initial WERGS sample whose radio emission might be
dominated by the star-formation of host galaxies, and not by the AGN, we set a lower-limit to radio-luminosity of $\lfirst > 10^{24}$~$\whz$, 
which is equivalent to the radio-emission
from the host galaxies with a star-formation rate of $\mathrm{SFR} \approx 250~M_\odot$~yr$^{-1}$ \citep{con92,con13}. This cut is also supported by a steep decline of radio luminosity function of starburst galaxies 
above $\lfirst = 10^{23}$~$\whz$
\citep[e.g.,][]{kim11,tad16,pad16}. This criterion reduces the sample size to 3147.


Next, we limit our sample to compact sources in the radio band.
This is important in order to reduce false optical identification by mismatching the optical
sources with the locations of spatially 
extended radio-lobe emission.
\cite{yam18} discussed this and categorized radio compact sources using the ratio of the total integrated radio flux density to the peak radio flux density $f_\mathrm{int}/f_\mathrm{peak}$.
They treated the source as compact if the ratio fulfills 
either of two following equations,
\begin{align}
f_\mathrm{int}/f_\mathrm{peak} &< 1 + 6.5 \times (f_\mathrm{peak}/\mathrm{rms})^{-1}\\
\log (f_\mathrm{int}/f_\mathrm{peak}) &< 0.1,
\end{align}
where $f_\mathrm{peak}/\mathrm{rms}$ is the S/N, and rms is a local rms noise in the FIRST catalog, and the above two equations are obtained from the study by \cite{sch07} \citep[see also the original criterion: ][]{ive02}.
We set an additional conservative requirement that there shall be
no FIRST source in the surrounding 1~arcmin (corresponding to $\lesssim 500$~kpc at $z\sim1$)
to select the isolated radio core emission and avoid coincident matching with radio lobes of other nearby FIRST sources.
This reduces the sample to 2583 sources.

A further ten sources were removed in crowded regions of the HSC footprint due to multiple HSC detections within $<1$~arcsec of the HSC counterpart, bringing the sample down to 2573 sources.

Finally, we restricted the sample to sources with spectroscopic redshift
(spec-$z$; $\specz$) or reliable photometric redshift (photo-$z$; $\photz$) values. 
The spec-$z$ were obtained
from SDSS DR12 \citep{ala15}, 
the GAMA project DR2 \citep{dri11,lis15},
and WiggleZ Dark Energy Survey project DR1
\citep{dri10}.
In this study, we limit the sample to the spec-$z$ range $0.3 < \specz < 1.6$ to cover the same redshift range as
the photo-$z$ sample.
For the sources without spec-$z$, 
we utilized the photo-$z$ estimated using
the \verb|Mizuki| SED-fitting code
which is one of the standard photo-$z$ packages for the HSC-SSP survey.
The method utilizes the photometries of the five HSC-SSP bands
\citep[see][for more details]{tan15,tan18} for the SED fitting.
\cite{yam18} discussed that the HSC-SSP photo-$z$ derived by \verb|MIZUKI| is reliable for $\photz<1.6$ based on comparison with spectroscopic redshift in
the COSMOS field.
In addition, the redshift of sources with $\photz<0.3$ 
are sometimes erroneous because they lack
the Balmer break tracer in the HSC optical bands.
Therefore, we limit the sample to $0.3 < \photz < 1.6$.
We further imposed requirements
based on the reliable photo-$z$ fitting quality (reduced $\chi^2 < 3$, $\sigma/\photz<0.2$, and
$\sigma/(1+\photz)<0.1$),
following \cite{tob19a}.
The resulting WERGS sample contains 
1770 sources.
Given that our sample largely relies on photo-$z$ results, we further discussed how possible erroneous photo-$z$ might affect our main results in Appendix~\ref{sec:appendix}.

\setlength{\topmargin}{3.0cm}
\floattable
\renewcommand{\arraystretch}{1.1}
\begin{deluxetable}{cccccccc}
\thispagestyle{empty}
  \tablecaption{Summary of the sample selection cut in this study}\label{tab:Equation}
  \tablehead{
      \colhead{(1)} & \colhead{(2)} & \colhead{(3)} 
      \\
       \colhead{Selection} & \colhead{No. of WERGS sources} & \colhead{No. of \ERGs} 
     }
    \startdata
\thispagestyle{empty}
Parent WERGS sample of \cite{yam18} & 3579 & 273 & \\
$\lfirst > 10^{24}$~W~Hz$^{-1}$ & 3147 & 270 & \\
compact radio sources & 2792 & 208 & \\
No nearby multiple FIRST sources in $<1$~arcmin & 2583 & 192 & \\
No nearby multiple HSC sources in $<1$~arcsec & 2573 & 187 & \\
Redshift range at $0.3<z<1.6$ & 2286 & 145 & \\
Good photo-$z$ quality & 1770 & 79 & \\
IR available region by \cite{tob19a} & 1055 & 39 & \\
$\qir < 1.68$ (final sample) & 1026 & 39 & \\
    \enddata
    \thispagestyle{empty}
\tablenotetext{}{
Notes.--- The number of sources after the each selection cut 
staring from the parent sample to the final one used in this study. 
The details are summarized at Section~\ref{sec:hsc} and \ref{sec:tob19}.
}\label{tab:selection}
\end{deluxetable}
\setlength{\topmargin}{0in}

\subsection{WERGS IR catalog}\label{sec:tob19}

To study the AGN and host galaxy properties in WERGS, 
\cite{tob19a} compiled optical, 
near-IR, mid-IR, far-IR, and radio data for the WERGS sample.
\cite{tob19a} performed spectral energy distribution (SED) fitting with \verb|CIGALE| \citep{boq19},
and inferred physical properties,
including the IR luminosity contributed
from AGN (IR AGN luminosity; $\liragn$), 
dust extinction corrected stellar-mass ($\mstar$), and SFR estimated from the decomposed host galaxy IR emission\footnote{The catalog containing physical parameter of \cite{tob19a} is available at
\url{http://vizier.u-strasbg.fr/viz-bin/VizieR?-source=J/ApJS/243/15}}.
For the estimation of $\liragn$, \cite{tob19a}
considered the contamination of Synchrotron radiation
in the IR bands by extrapolating the power-law from the radio bands. Thus, $\liragn$ is obtained purely
from the dust emission heated by AGN.
The bolometric AGN luminosity is estimated by using the conversion 
$\lbol \simeq 3 \times \liragn$ \citep{del14,ina18}.
Because \cite{tob19a} restricted themselves to objects
in an area of $\sim 95$~deg$^2$, in which mutli-wavelength 
information from $u$-band to far-IR was available,
our sample is reduced to 1055 sources.

 Some of the selected sources have intensive star formation rates, reaching $\mathrm{SFR} \approx 250 \msun$~yr$^{-1}$, so we applied an additional cut to remove possible star-formation dominated radio galaxies using the ratio of IR and radio luminosity \citep[$\qir$; ][]{hel85,ivi10} defined by
\begin{equation}
\qir = \log \left( \frac{L_\mathrm{IR} / 3.75\times 10^{12}}{\lfirst}  \right),
\end{equation}
where $L_\mathrm{IR}$ is the total IR luminosity in units of W derived from \verb|CIGALE| and $3.7\times10^{12}$ is the frequency in Hz corresponding to 80~$\mu$m, which makes $\qir$ a dimensionless quantity. We set $\qir < 1.68$ \citep{del13} to select radio excess (meaning jet-emission dominated) sources.
As a result, our final sample reduces to \Nwergstot~sources, and 
the resulting final number of ERGs fulfilling $\log \R > 4$ is \Nergs\ sources, and they are shown in orange stars in Figure~\ref{fig:R_vs_gmag}.
We refer to the remaining \Nwergsother\ WERGS sources with $1<\log \R <4$ as ``normal radio galaxies (\NRGs)'' (shown with blue circles in Figure~\ref{fig:R_vs_gmag}) hereafter.

Table~\ref{tab:selection} shows the number of sources for the full WERGS sample
and the \ERGs\ after each selection cut. 
\ERGs\ suffers the photo-$z$ quality cut more than the full WERGS sample because \ERGs\ contain
a relatively large fraction of optically faintest sources with $i_\mathrm{AB}>25$, which sometimes makes the reliable photo-$z$ estimation difficult.
The IR selection cut is another main factor
in reducing the number of sources. This is mainly because the region available for the IR data is limited in the  area of $\sim95$~deg$^2$ out of the 154~deg$^2$ \citep{tob19a}.
Note that this IR selection
reduces the similar fraction of sources for the full WERGS sample (60\%) and \ERGs\ (55\%).

\section{Results}\label{sec:analysis}

We summarize the properties of the obtained
\Nergs~\ERGs\ with IR detections.
First, we show the basic differences between \ERGs\ and \NRGs.
Then, we show the properties of \ERGs\ on SMBHs and host galaxies.

\subsection{Basic Sample Properties}

\subsubsection{$\log \R$ vs. $g$-band magnitude}\label{sec:R_vs_gmag}

Figure~\ref{fig:R_vs_gmag} shows the distribution of the radio-loudness parameter $\R$ as a function
of the rest-frame $g$-band magnitude
($g_\mathrm{AB}$). 
The vertical dashed-line at $g_\mathrm{AB} =22$
indicates the SDSS magnitude limit \citep[e.g.,][]{ive02}. 
Since most \ERGs\ are fainter than the SDSS limit, Subaru/HSC has enabled us to investigate this unique population for the first time.
Figure~\ref{fig:R_vs_gmag} also shows that most \ERGs\ are very faint in the optical, 
with median magnitudes of
$\langle g_\mathrm{AB} \rangle=24.5$.
The current 8~m class telescopes 
are sufficiently sensitive to obtain spec-$z$ for the bright end
of the \ERGs\ sample down to $g_\mathrm{AB} \approx 24.0$, 
and with 30~m class telescopes, spec-$z$ can be obtained for sources down to $g_\mathrm{AB} \approx 26.0$.

\begin{figure*}
\begin{center}
\includegraphics[width=0.48\textwidth]{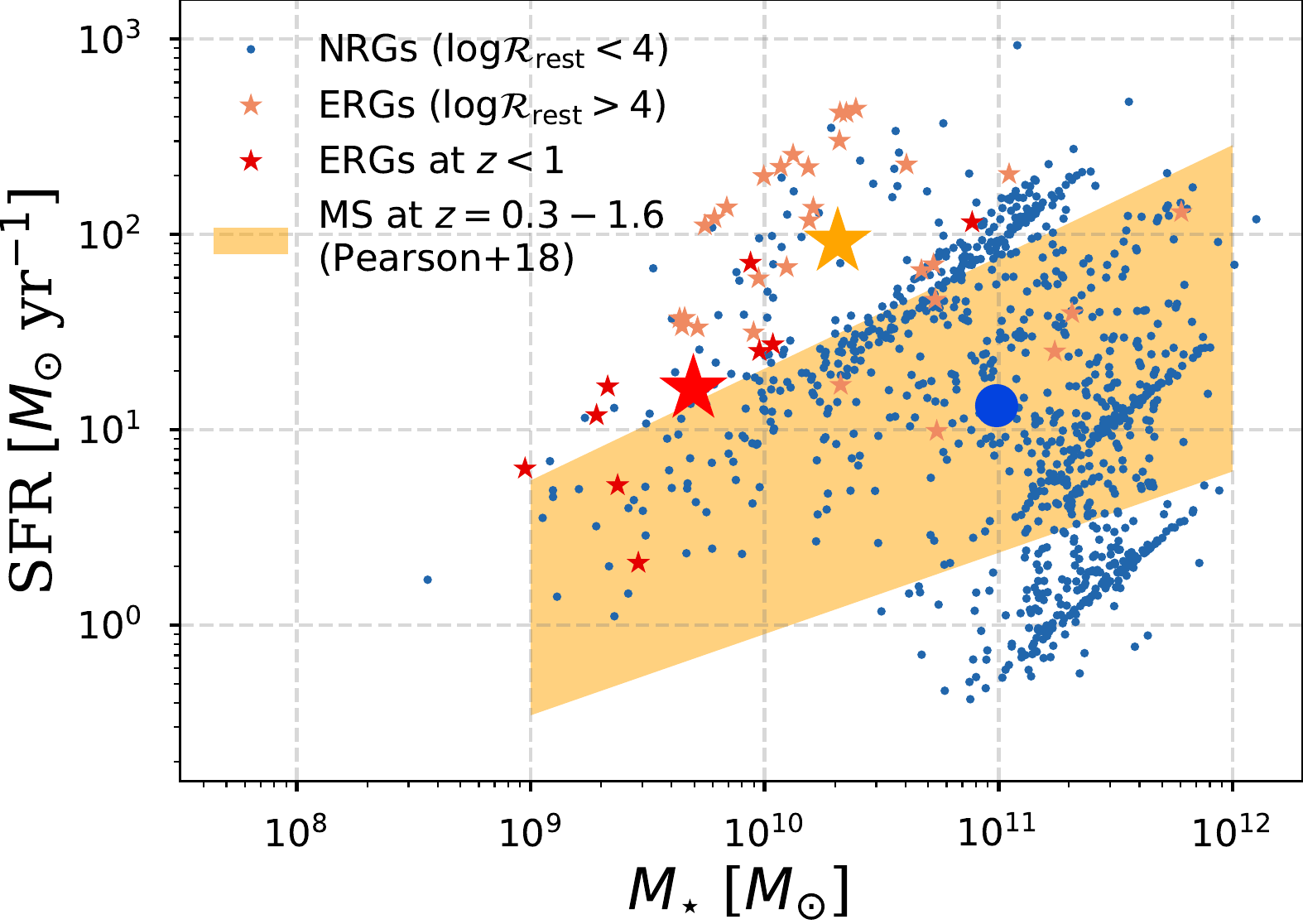}~
\includegraphics[width=0.48\textwidth]{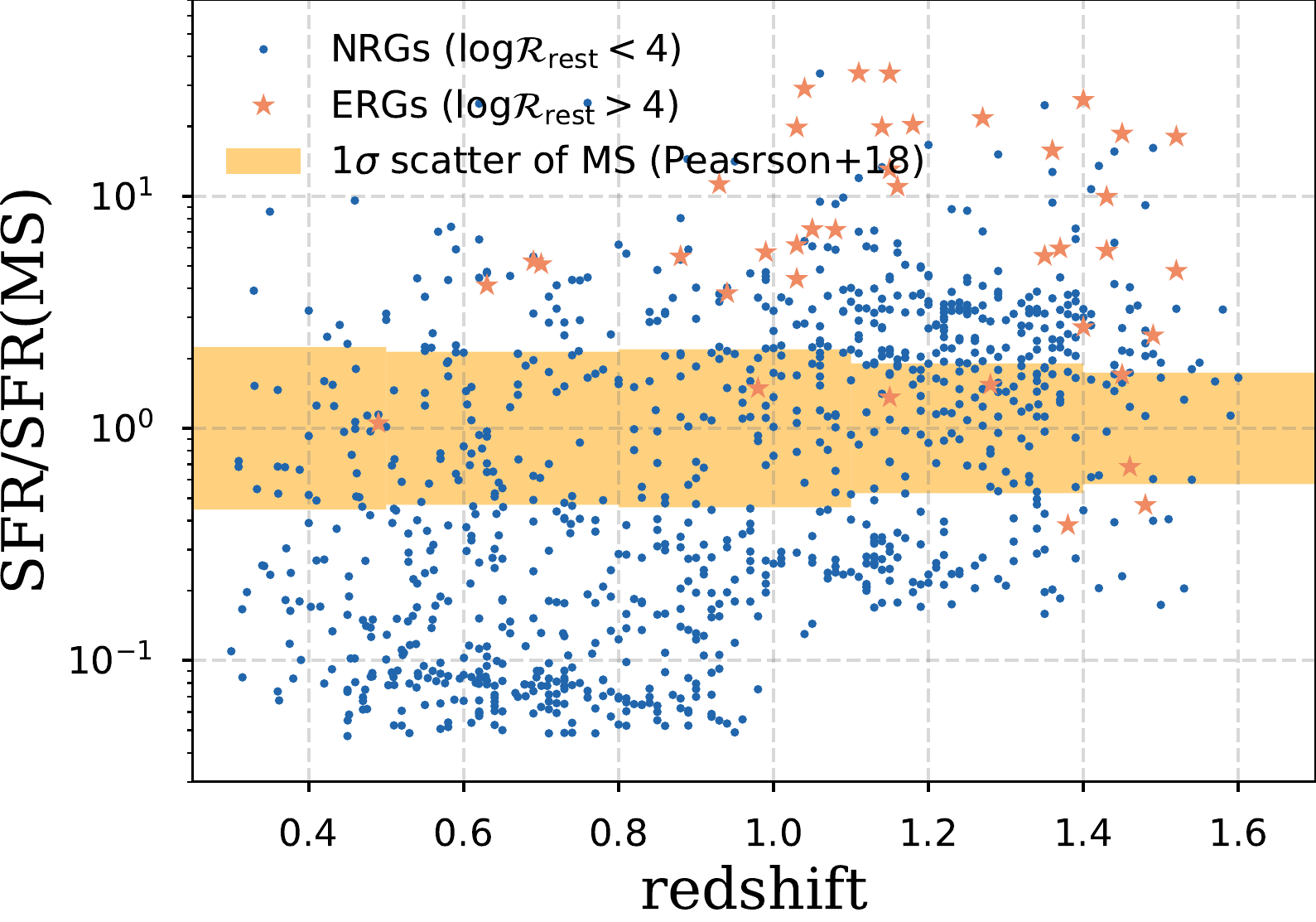}
\caption{
(Left) The relation between SFR and
stellar-mass ($\mstar$) of the WERGS sample. The both values are obtained from the SED fitting by \cite{tob19a}.
The blue circles represent the \NRGs. The orange and red stars represent \ERGs\ at $z>1$ and $z<1$, respectively.
The large points show the average of each sample.
The shaded area is the expected main-sequence region from $z=0.3$ (minimum) to $z=1.6$ (maximum) with the scatter of $\sigma=0.35$ at $z=0.3$ and $\sigma=0.24$ at $z=1.6$, which is obtained from \cite{pea18}. The redshift evolution of the relation is summarized in Figure~\ref{fig:highz}.
(Right) Redshift dependence of the ratio SFR/SFR(MS) of the WERGS sample.
The 1$\sigma$ scatter is shown with the orange shaded area, which is also obtained
from \cite{pea18}.
}\label{fig:SFRMstar}
\end{center}
\end{figure*}

\subsubsection{$L$--$z$ plane}\label{sec:Lzplane}

Figure~\ref{fig:L_vs_z} shows the radio and IR AGN luminosities of our sample as a function of redshift.
The 1.4~GHz radio luminosity ($\lfirst$~W~Hz$^{-1}$) 
is taken from \cite{yam18} or \cite{tob19a}, in which
the integrated flux density ($f_\mathrm{int}$) obtained by the VLA/FIRST final catalog \citep{hel15} is used,
and $k$-correction is applied by assuming a power-law radio spectrum with $f_\nu \propto \nu^{\alpha}$,
where $\alpha$ is estimated with $\alpha = \log (f_\mathrm{1.4GHz}/f_\mathrm{150MHz})/\log (\nu_\mathrm{1.4GHz}/\nu_\mathrm{150MHz})$ for the objects having TGSS 150~MHz data \citep{int17}, and $\alpha=-0.7$ for all others \citep[e.g.,][]{con92}.

As shown in the left panel of Figure~\ref{fig:L_vs_z},
\ERGs\ show relatively high radio luminosities with a median of $ \langle \log (L_\mathrm{1.4GHz}/\mathrm{W~Hz{^{-1}}}) \rangle = 26.3$, 
which is one order of magnitude higher than that of 
the \NRGs\ ($ \langle \log (L_\mathrm{1.4GHz}/\mathrm{W~Hz{^{-1}}}) \rangle = 25.1$).
On the other hand, in terms of the IR AGN luminosity shown in the
right panel, there is no clear difference between \ERGs\ and the \NRGs\:
 $\langle \log (\liragnunit) \rangle = 44.9$ for \ERGs\ and $\langle \log (\liragnunit) \rangle = 44.6$ for the \NRGs.
We therefore assume \ERGs\ to be intrinsically radio-louder sources than \NRGs, 
and we will discuss this in Section~\ref{sec:highR}.

\begin{figure}
\begin{center}
\includegraphics[width=0.48\textwidth]{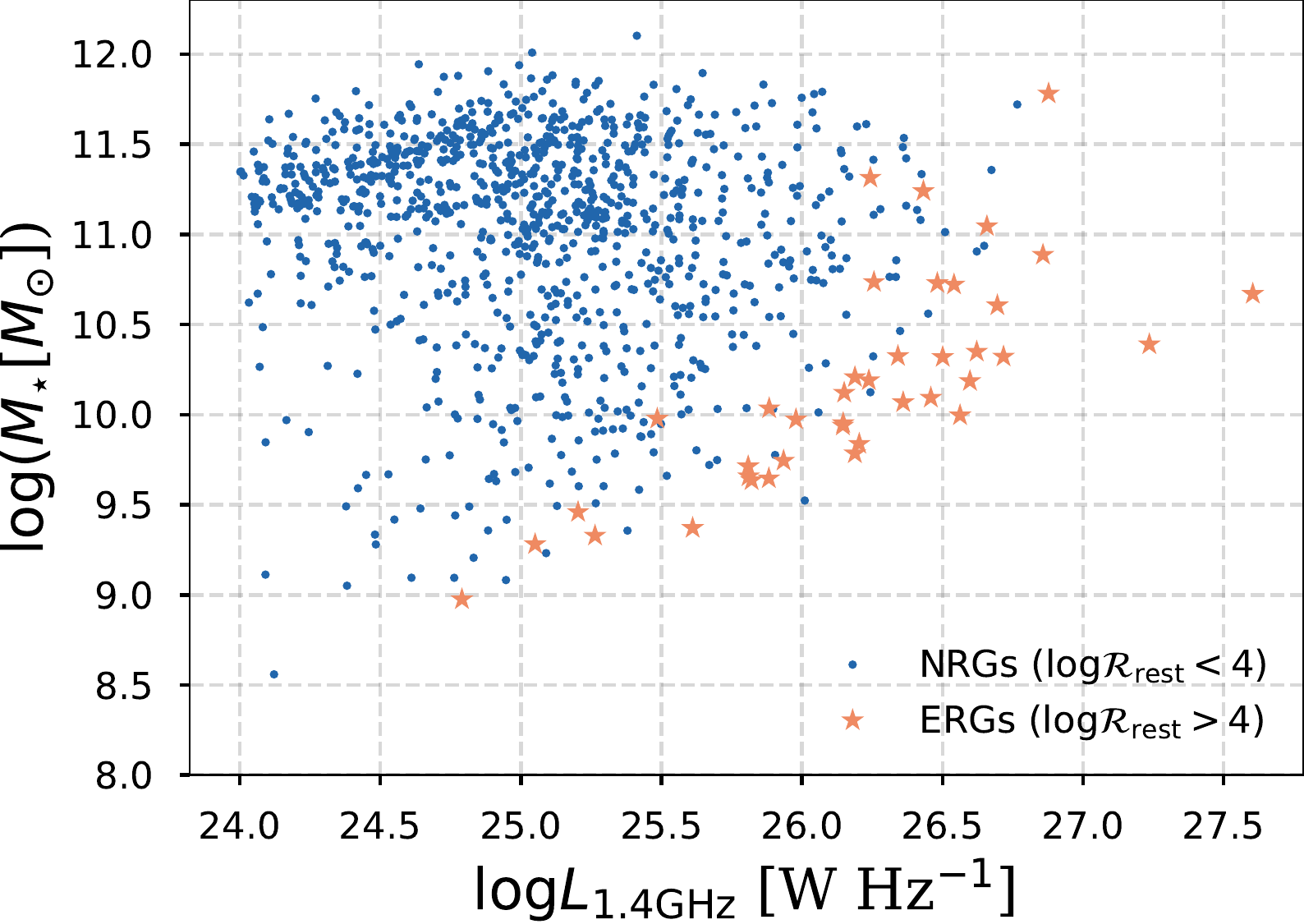}
\caption{
The stellar-mass ($\mstar$) as a function of
the radio luminosity $\log \lfirst$. The points are the same as in Figure~1.
}\label{fig:Mstar_vs_L1p4}
\end{center}
\end{figure}

\subsection{Host Galaxy Properties}

The left panel of Figure~\ref{fig:SFRMstar} shows the distribution of our RGs in the
SFR and $\mstar$ plane, where both values are obtained from the optical+IR SED fitting with \verb|CIGALE| \citep{tob19a}.
It is known that most star-forming galaxies follow the main-sequence (MS)
whose normalization
moves upward with redshift, as galaxies are more gas-rich at higher $z$
\citep[e.g.,][]{bri04,noe07, elb11, whi12, sch15b}.
Above the MS, galaxies are referred to as starburst galaxies, producing stars more efficiently than regular star-forming galaxies.
\cite{pea18} measured the SFR and $\mstar$ 
using multi-wavelength data from UV to far-IR
employing the \verb|CIGALE| code, and their
 MS is delineated with the shaded area in Figure~\ref{fig:SFRMstar}, whose bottom and top boundaries correspond to redshifts of $z=0.3$ and $z=1.6$
 with the scatter of $\sigma=0.35$ at $z=0.3$ and $\sigma=0.24$ at $z=1.6$, respectively.
 In this study, we define the sources as starburst galaxies if they are located above the yellow shaded MS area, which is roughly consistent with the criteria in previous studies \citep[e.g.,][]{elb11}.
 
 Considering that our sample spans a wide redshift range in $0.3 < z < 1.6$ and their rapid redshift evolution of MS, redshift dependence is a key factor to understand the relation in the left panel of Figure~\ref{fig:SFRMstar}. 
Therefore we also calculate the ratios between the observed SFRs and those predicted from the MS relations at a given stellar-mass, that is SFR/SFR(MS),
and its redshift dependence is shown in
the right panel of Figure~\ref{fig:SFRMstar}.
 We also compile how the SFR and $\mstar$ relation evolves with redshift in Appendix~\ref{sec:appendix_zdep} and Figure~\ref{fig:highz} (top panels).
 
Figure~\ref{fig:SFRMstar} shows several sequences of sources, notably at around
sSFR$\sim10^{-10.5}$, $\sim 10^{-9}$, and $\sim 10^{-8}$~yr$^{-1}$, and the sources locate scarcely between the sequences. This does not originate from the properties of nature, but more mainly from
the complications of the limited bin numbers of the SED fitting parameters of star-forming history (SFH). In our case, the most notable contribution is considered to be $f_\mathrm{burst}$; mass fraction of the burst population. \cite{tob19a} set a limited parameter binning of $f_\mathrm{burst} = 0.001, 0.1$, and $0.3$, which is roughly corresponding to the bindings at sSFR$\sim10^{-10.5}, 10^{-9},$ and $10^{-8}$~yr$^{-1}$.
 This banding might disappear once we follow the same SFH parameter binning of \cite{pea18}, which was impossible for our studies because of the additional parameters of AGN and radio components
 which increase the computational time exponentially.
 Therefore instead we assume that SFRs in this study have large uncertainty of $\sim0.5$~dex. Please see the more detailed discussion on the choice of the SFH parameters and their effects on the SFR estimations \citep[e.g.,][]{sch15a,cie17,pea18}.
 
Although our SFR estimation by \cite{tob19a} might harbor large uncertainties as discussed above, Figure~\ref{fig:SFRMstar} shows three important suggestions.
One is that most \ERGs\
are distributed above the MS by a factor of $\sim10$ (see the right panel), 
suggesting that most \ERGs\ are in starburst mode and they contain a copious amount of gas.
This is indirect evidence for the availability 
of cold gas supply fueling both star-formation, and the central SMBHs.
The specific star-formation rate defined by 
$\sSFR = \SFR/\mstar$ reaches 
$\mathrm{sSFR}\sim10^{-8}$~yr$^{-1}$ for ERGs, suggesting that ERGs are in a rapid stellar-mass assembly phase with mass doubling times of only $\sim100$~Myr.
This value is also comparable to one starburst duration \citep[e.g.,][]{mcq10} and thus most of the stellar component is expected to be dominated by 
a very young population.

Second, the onset of AGN feedback on the host galaxies has apparently not (yet) happened to most \ERGs\ considering their location above the MS.
This indicates a different picture from the conventional local RGs, which show a preference of massive host galaxies 
($\mstar > 10^{11} \msun$) and fall below the MS because their strong AGN feedback quenches star formation \citep[e.g.,][]{sey07}.
Note that our sample requires IR detections for the \verb|CIGALE| SED fitting to derive stellar-mass and SFR, 
therefore the WERGS not detected in IR do not
appear in Figure~\ref{fig:SFRMstar}.
In addition to that, our sample requires radio compact morphologies, which removes a large population of local, radio-extended RGs.
These two selection criteria may introduce a selection bias,
since it is possible that these 
IR non-detection or radio extended sources are clustered below the SF main-sequence.
It is nonetheless intriguing that ERGs are located at the top edge of the distribution within the WERGS sample with IR detections.

The third point is that, if we limit ourselves to the sources with $z<1$ (red star points), 
most of these are clustered at the low-$\mstar$ end, typically at $\log \mstar/\msun < 10$. 
This suggests that the low-$z$ subsample of \ERGs\ is potentially important with regard to the study of 
BH seeds.
Considering that all of our sources fulfill (1) $\lfirst > 10^{24}$~$\whz$, that is, 
they lie above the threshold where their radio luminosity could be produced by star formation,
as discussed in Section~\ref{sec:hsc}, and (2) the radio-excess selection with $\qir < 1.68$, their radio emission can only 
be achieved by the AGN jet, requiring the existence of a SMBH at the center.
Therefore, \ERGs\ contain massive BH candidates residing in low-mass galaxies.

Figure~\ref{fig:Mstar_vs_L1p4} shows the distribution of \ERGs\ and
\NRGs\ in the $\mstar$ and $\lfirst$ plane,
and illustrates the third point above from a different aspect.
\ERGs\ populate regions of comparatively small stellar-mass and high $\lfirst$, showing that
 higher $\R$ sources tend to have smaller $\mstar$ as well as higher $\lfirst$.
The combined results from above and from Figure~\ref{fig:R_vs_gmag} suggest that smaller $\mstar$ sources
tend to observed as optically fainter sources,
and are therefore observed as high $\R$ sources. 
We will discuss this point later in Section~\ref{sec:highR}.

\begin{figure*}
\begin{center}
\includegraphics[width=0.95\textwidth]{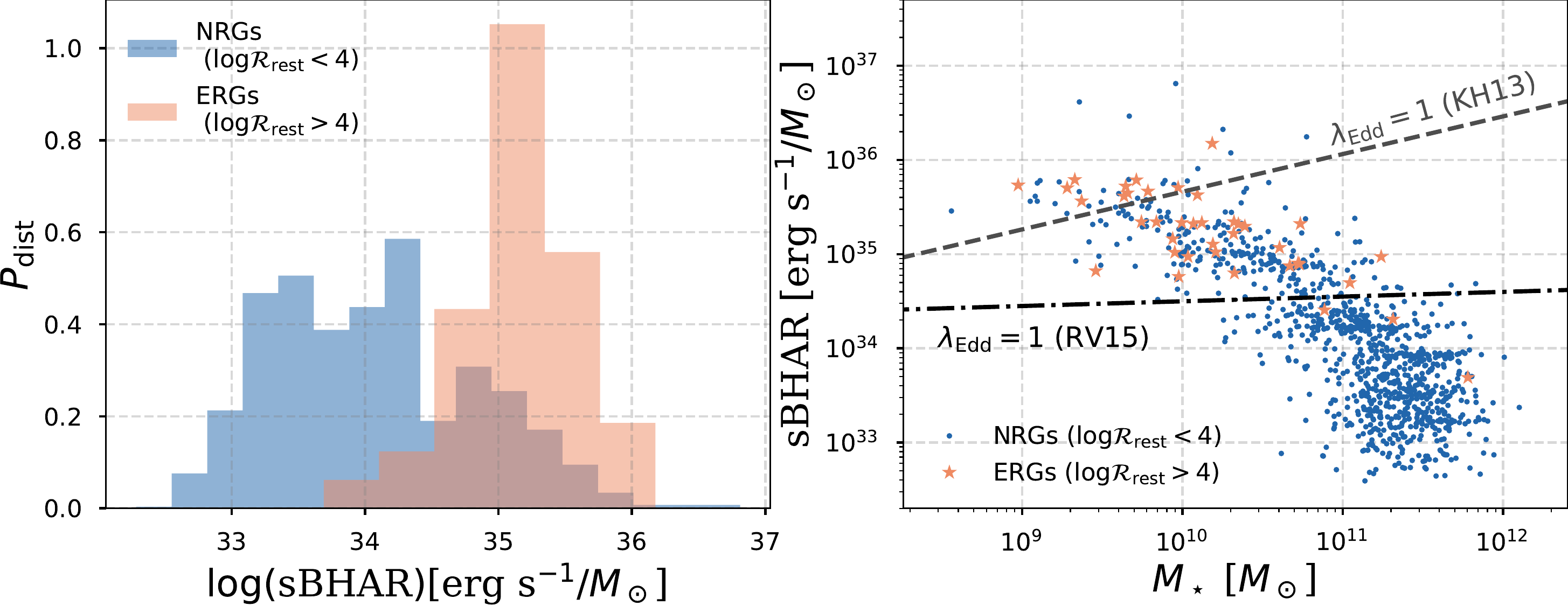}
\caption{
sBHAR properties for \ERGs\
and \NRGs\ in this study. 
The colors and symbols are same as in Figure~\ref{fig:R_vs_gmag}.
(Left) the distribution of sBHAR. 
The median of sBHAR is 
$\langle \log (\mathrm{sBHAR}/\sBHARunit) \rangle = 35.3$ for \ERGs\ and $\langle \log (\mathrm{sBHAR}/\sBHARunit) \rangle = 34.0$
for \NRGs.
(Right) The relation between sBHAR and $\mstar$.
The two straight lines are the expected Eddington-limits
 $\lambda_\mathrm{Edd}=1$ using Equation~(\ref{eq:sBHAR_KH13}; gray
dashed line) and using Equation~(\ref{eq:sBHAR_RV15}; black dot-dashed line). The redshift evolution of the relation is summarized in the middle panels of Figure~\ref{fig:highz}.
}\label{fig:sBHARvsMstar}
\end{center}
\end{figure*}

\subsection{SMBH Properties: Relation between sBHAR and $\mstar$}\label{sec:sBHAR}

Considering that most \ERGs\ are very faint ($\left< g_\mathrm{AB} \right> =24.5$) in optical, 
as shown in Figure~\ref{fig:R_vs_gmag} and Section~\ref{sec:R_vs_gmag}, 
obtaining the BH virial mass from spectroscopy \citep[e.g.,][]{mcl04} is time-consuming and difficult even if they are type-1 AGN.
This means that the direct measurement of the Eddington ratio ($\lambdaedd \equiv \lbol / \ledd$,
where $\ledd$ is Eddington luminosity $\ledd \simeq 1.26 \times 10^{38} (\mbh/\msun)$~erg~s$^{-1}$),
is difficult at this stage.
Instead, we investigate the SMBH properties through the specific black hole accretion rate (hereafter, sBHAR), which may be considered a proxy for the Eddington ratio. The sBHAR is conventionally defined as $\mathrm{sBHAR}=\lbol/M_\star$~$\sBHARunit$ \citep[e.g.,][]{mul12}.
The left panel of Figure~\ref{fig:sBHARvsMstar} shows the distribution of sBHAR
for \ERGs\ and \NRGs.
The median sBHAR of \ERGs\ is 
$\langle \log (\mathrm{sBHAR}/\sBHARunit) \rangle = 35.3$,
which is an order of magnitude higher than
that of the \NRGs\ with $\langle \log (\mathrm{sBHAR}/\sBHARunit) \rangle = 34.0$.
Based on \cite{bes12}, the boundary of radiatively inefficient and efficient AGN
is at an Eddington ratio of $\log \lambdaedd \sim -2$, which corresponds to $\log (\mathrm{sBHAR}/\sBHARunit)=32$--$33$. 
 Thus, almost all of our sources are considered to be radiatively efficient AGN.

This difference in sBHAR between
the two subgroups is more pronounced in the right panel of Figure~\ref{fig:sBHARvsMstar},
showing sBHAR as a function of $\mstar$.
To illustrate the dependence of $M_\star$ and $\lambda_\mathrm{Edd}$ on sBHAR more clearly,
we use local scaling relations between $M_\mathrm{BH}$ and $\mstar$.
Previous studies assumed
 a constant ratio of $\mbh/M_\mathrm{bulge} = (1.5-2.0)\times 10^{-3}$ \citep{mar03,har04}
and $M_\mathrm{bulge} \approx \mstar$ for simplicity \citep{air12,mul12,del18,air19}.
In this study, 
we apply stellar-mass dependent values for $\mbh$ with $\mbh \propto M_\star^\beta$, where 
$\beta=1.4$ \citep[KH13;][]{kor13} and $\beta=1.05$ \citep[RV15;][]{rei15}.
With this relation, sBHAR can be written as a function 
of $\lambda_\mathrm{Edd}$ and $\mstar$ by
\begin{align}
\mathrm{sBHAR (KH13)}= 4.6\times 10^{35} \lambda_\mathrm{Edd} \left( \mstar / 10^{10}~\msun \right)^{0.4}, \label{eq:sBHAR_KH13}\\
\mathrm{sBHAR (RV15)}= 3.2\times 10^{34} \lambda_\mathrm{Edd} \left( \mstar / 10^{10}~\msun \right)^{0.05}. \label{eq:sBHAR_RV15}
\end{align} 

The right panel of Figure~\ref{fig:sBHARvsMstar} exhibits that \ERGs\ are more clustered
in the high Eddington ratio regime.
The expected median Eddington ratio by using Eq.~(\ref{eq:sBHAR_KH13}) is 
$\left< \log \lambdaedd \right> \approx -0.4$.
Some sources appear to exceed the Eddington limit.
\NRGs\ cover a broader range in sBHAR, and some massive galaxies with $\mstar > 10^{11} \msun$ fall into the radiatively in-efficient regime
with $\lambdaedd < -2$.

We caution that the above discussion does not take into account the redshift evolution of the
$\mbh$--$\mstar$ relation, 
which has been suggested in some studies \citep[e.g.,][]{woo08,mer10,din20}.
In this case, the estimated $\mbh$ becomes larger at a given $\mstar$, and the resulting Eddington ratio becomes smaller.
Assuming the evolutional trend of \cite{mer10}, with $\mbh/\mstar (z) \propto (1+z)^{0.68}$, and 
using a median redshift for the \ERGs\ of $\langle z \rangle = 1.1$, the normalization of sBHAR
becomes slightly smaller by a factor of 1.6.
As shown in Figure~\ref{fig:sBHARvsMstar}, this difference does not change our overall result.
Although there are other significant uncertainties and possible systematics of the expected $\lambda_\mathrm{Edd}$ values,
under the basic assumptions of Equations~(\ref{eq:sBHAR_KH13}) or (\ref{eq:sBHAR_RV15}), 
our results suggest that higher $\R$ sources have higher sBHAR, and \ERGs\ might contain sources achieving super-Eddington accretion.
We will discuss this point in Section~\ref{sec:highR}.

In summary, our results indicate that \ERGs\ 
exhibit high sBHAR with the expected Eddington ratio of $\left< \log \lambdaedd \right> \approx -0.4$, and some \ERGs\ may be experiencing a super-Eddington phase. 
In addition, \ERGs\ are likely starburst galaxies and 
their stellar-mass is relatively lower than \NRGs. \ERGs\ appear to be in a phase preceding the onset of AGN feedback. 
If we limit the sample to $z<1$, most \ERGs\ are low-mass galaxies with $\mstar <10^{10} \msun$. Therefore, lower-$z$ \ERGs\ are also candidates of massive, rapidly growing BHs.

\section{Discussion}\label{sec:discussion}

\begin{figure*}
\begin{center}
\includegraphics[width=0.33\textwidth]{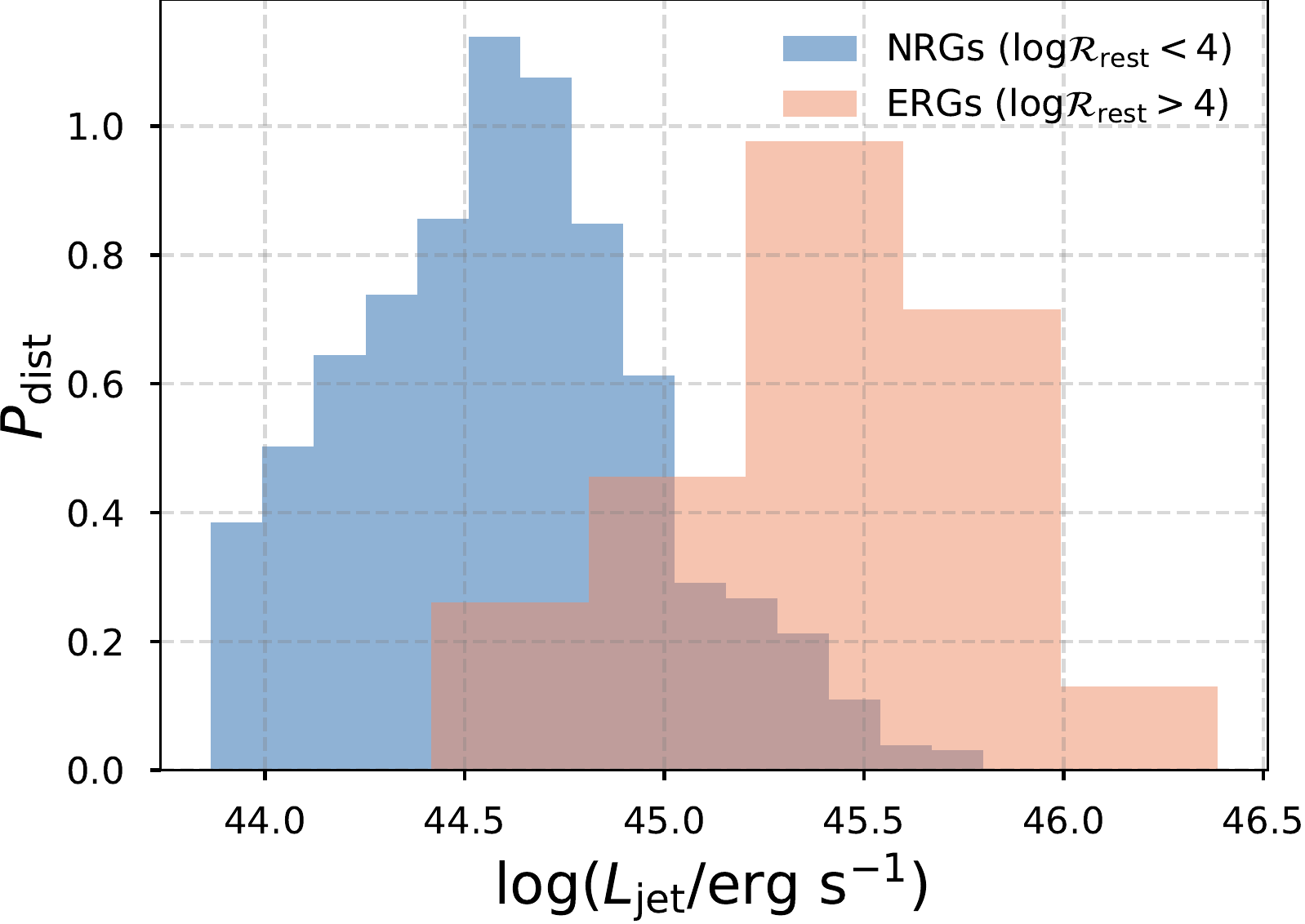}~
\includegraphics[width=0.33\textwidth]{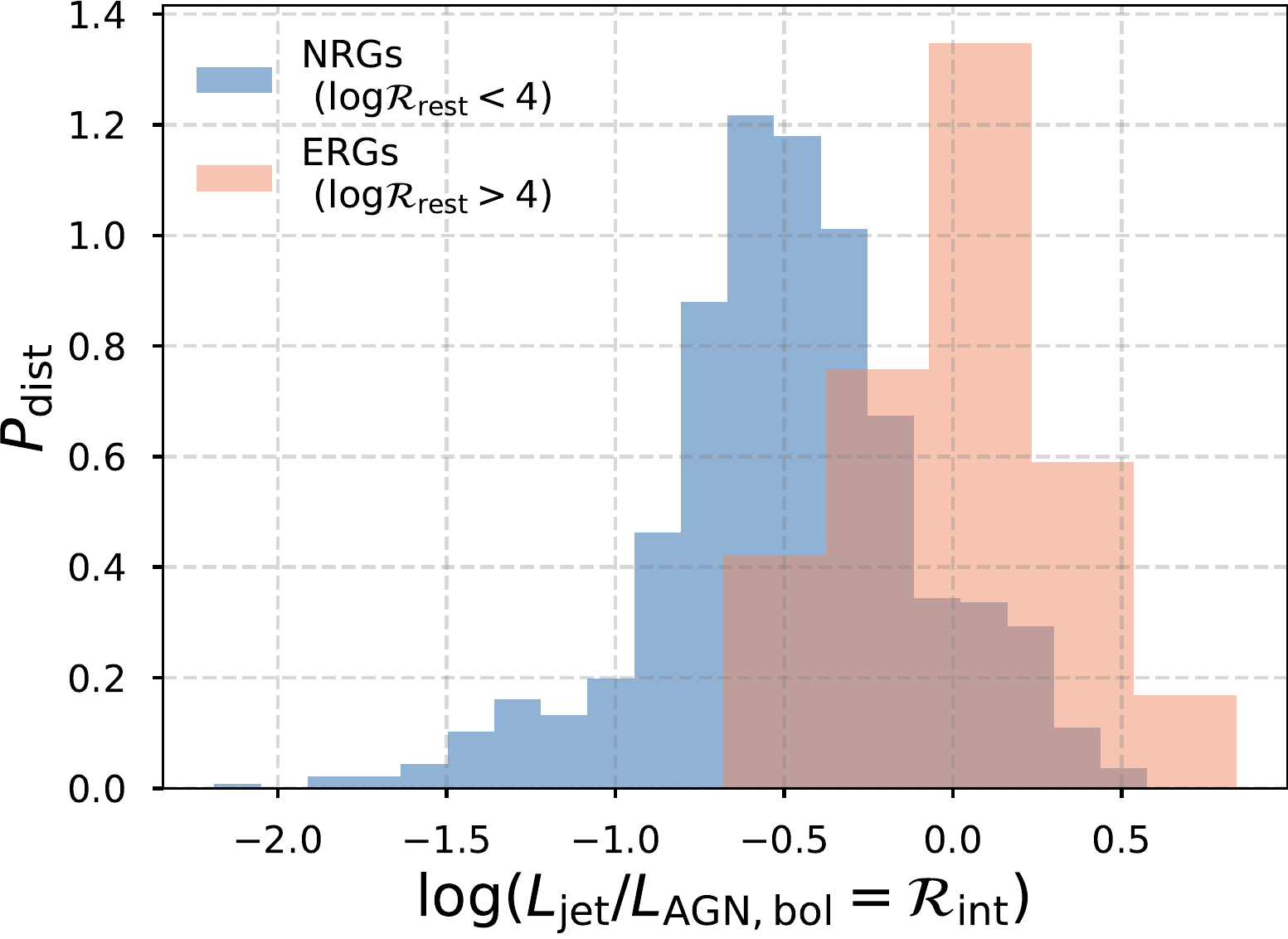}~
\includegraphics[width=0.33\textwidth]{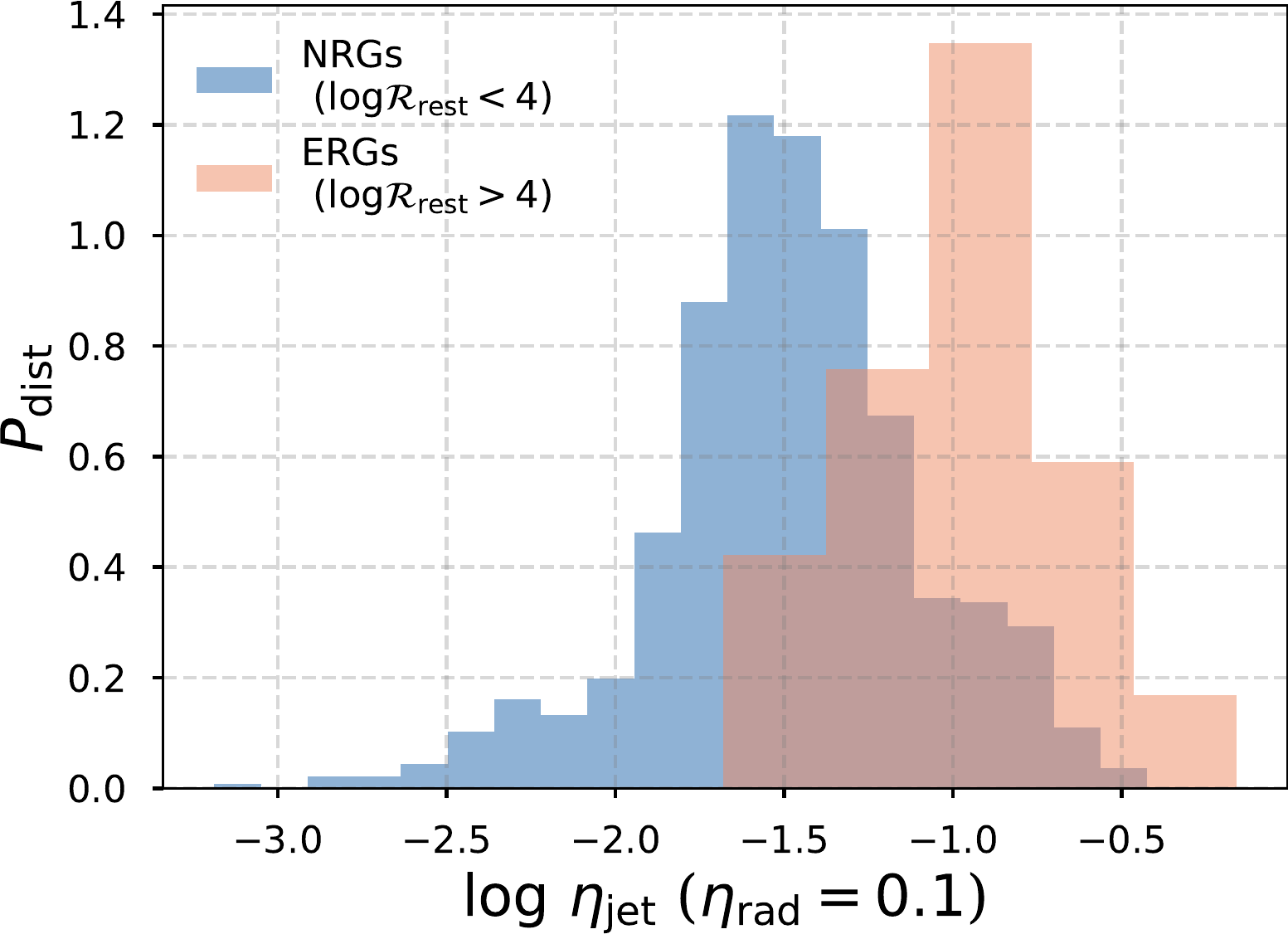}
\caption{
The distribution of $\ljetunit$ (left), $\ljet/\lbol$, an indicator of intrinsic $\Rint$ (middle),
 and $\etaj$ assuming the radiative efficiency of 0.1 (right).
 The color is same as in Figure~\ref{fig:R_vs_gmag}.
}\label{fig:hist_Rint}
\end{center}
\end{figure*}

\begin{figure*}
\begin{center}
\includegraphics[width=0.85\textwidth]{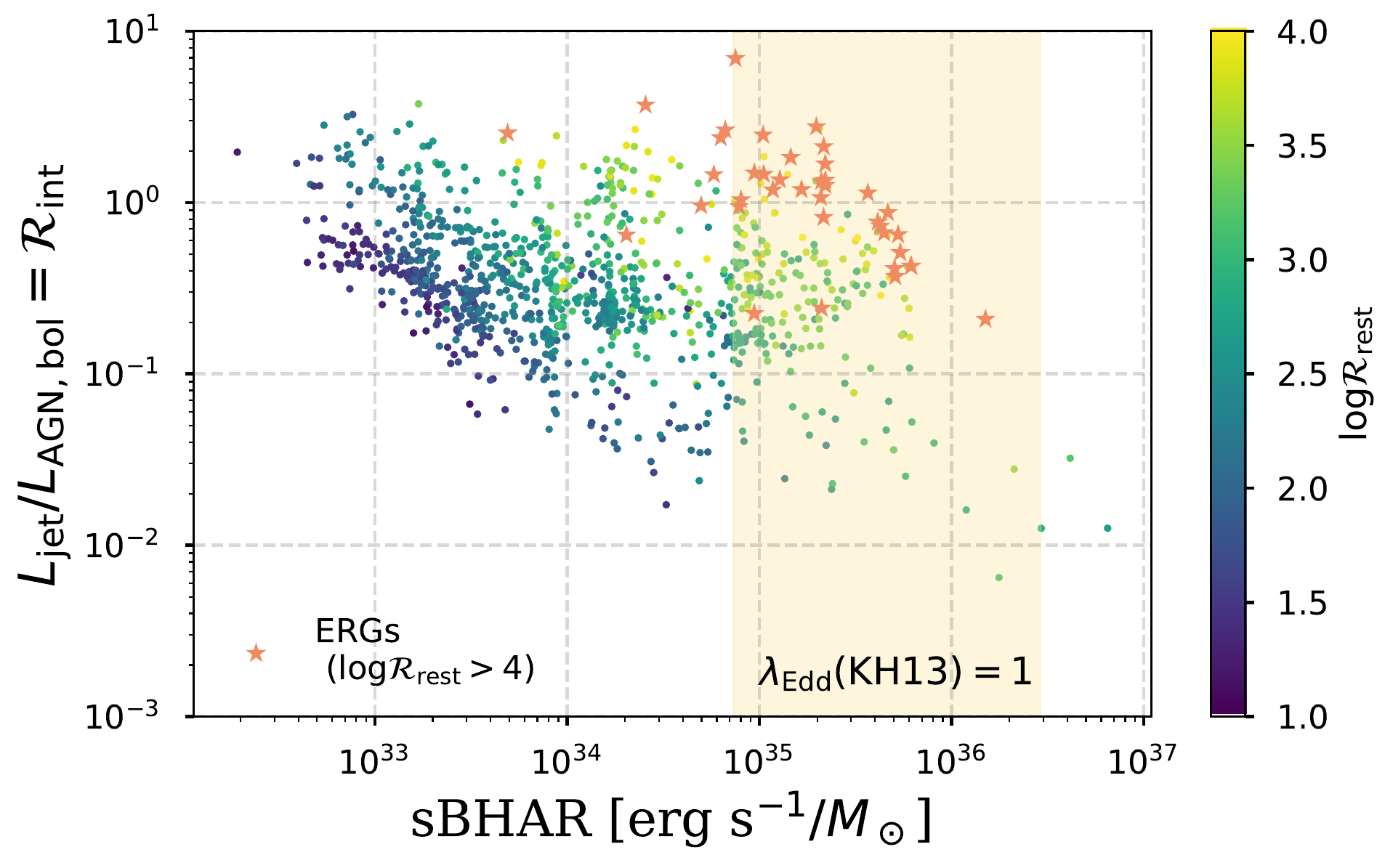}\\
\caption{
The relationship between $\ljet/\lbol = \Rint$ and sBHAR.
The yellow shaded area represents the corresponding $\lambdaedd=1$ region
assuming that $\mbh$ is based on the $\mbh$--$\mstar$ relation by \cite{kor13}.
 The symbols are same as those in Figure~\ref{fig:R_vs_gmag}, but
 other WERGS sources are shown with color gradation based on $\R$
 in the range of $1<\log \R < 4$.
The redshift evolution of the relation is summarized in the bottom panels of Figure~\ref{fig:highz}.
}\label{fig:R_vs_sBHAR}
\end{center}
\end{figure*}

\subsection{What Does High $\R$ Mean?}\label{sec:highR}

Figures~\ref{fig:R_vs_gmag} and \ref{fig:L_vs_z} indicate that 
\ERGs\ or high $\R$ sources tend to trace
optically-faint, but radio-loud AGN.
In addition, Figure~\ref{fig:Mstar_vs_L1p4} shows that higher $\R$ tends to pick up relatively
lower stellar-mass population. This means that the selection of \ERGs\ using $\R$ has a preference of selecting smaller $\mstar$ and, therefore, $\R$
using optical bands does not seem to trace AGN accretion disk emission in the optical band.
This is not a new result, but it is
a long-standing issue surrounding the use of the 
optical radio-loudness $\R$: it is strongly affected by host galaxy contamination and/or the nuclear obscuration around the SMBHs \citep{ter03,ho08}.
This raises the question of whether \ERGs\ are intrinsically radio-loud or not.

To investigate this, 
we estimate the ``intrinsic'' radio-loudness $\Rint$;
the  energy  balance  between  the  jet  and  accretion-disk defined by $\Rint = \ljet/\lbol$,
which is frequency-independent, and also free from the contamination of host galaxy component, unlike $\R$ \citep[e.g.,][]{ter03}.
For the accretion disk luminosity,
we use the bolometric AGN luminosity $\lbol$ obtained from the IR AGN luminosity ($\liragn$), since $\liragn$ is derived from IR SED decomposition, which traces the re-radiation of the dust powered by AGN and is mostly insusceptible to absorption \citep{gan09,asm15,ich12,ich17a,ich19a}.

The total jet power of the AGN ($\ljet$) can be estimated from the
 radio luminosity of the compact core or from the total radio emission \citep{wil99,cav10,osu11}.
 We adopt the relation between $\ljet$ and $\lfirst$ of \cite{cav10},
 \begin{align}
 \ljet = 7.3 \times 10^{43}~(\lfirst/10^{24}~\whz)^{0.70}~\mathrm{erg}~\mathrm{s}^{-1} \label{eq:ljet_lfirst}.
 \end{align}
Note that the above equation is derived with the assumption that most radio sources have a spectral index of $\alpha=0.8$. This value is almost consistent with the median value of our WERGS sample; $\alpha=0.65$ for the sources with detections at both TGSS 150~MHz and VLA/FIRST 1.4~GHz, or the fiducial value of $\alpha=0.7$ for the sources whose radio band detection is only in one band (i.e., VLA/FIRST 1.4~GHz only) \citep{con92}.
 \cite{sha13} recently investigated the effect of the radio source size in the jet power estimation,
 and found the dependence on the source size
 to be $\propto D^{0.58}$, suggesting that
 the effect of the source size is small.
 Since our sample is selected only for radio-compact emission in VLA/FIRST 
 with a spatial resolution of $\sim 5$~arcsec
  (or $\lesssim 20$~kpc at $z\sim1$)
this again mitigates the size dependence; otherwise the sources are ultra-compact sources.

The left panel of Figure~\ref{fig:hist_Rint} shows the distribution of $\ljet$,
spanning $44.0<  \log (\ljetunit) < 46.4$,
and the median values are
$\left< \log \ljetunit \right> = 45.4$ for \ERGs\
and $\left< \log \ljetunit \right> = 44.6$ for \NRGs,
which is equivalent to the radio-loud quasar level \citep[e.g.,][]{bes05,ino17}.
This is naturally expected since we apply the simple conversion relation (Eq.~\ref{eq:ljet_lfirst}), all of our
sources are selected based on VLA/FIRST detection with the $>1$~mJy flux limit, and the radio luminosity range is similar to that of the FIRST-SDSS sample,
although they are located in relatively higher-$z$ (see Figure~\ref{fig:L_vs_z}). 

This high $\ljet$ is powerful enough to produce a radio jet with the size of $\gtrsim 10$~kpc 
that can disperse the interstellar medium \citep[e.g.,][]{nes17} and create cavities in the
galactic inster-stellar medium and larger scale intergalactic environment to possibly quench star-formation in host galaxies \citep[e.g.,][]{mcn07}.
Given the compact radio sizes of the sources in our sample ($<5$~arcsec or $\lesssim 20$~kpc at $z\sim1$)
and given the fact that the host galaxies are still in an active star-forming phase (Figure~\ref{fig:SFRMstar}),  
\ERGs\ and the WERGS sources above or on the SF main-sequence may still be in the phase before the onset of huge kinetic radio feedback.

The middle panel of Figure~\ref{fig:hist_Rint} shows the distribution of $\Rint$,
and the median values of the two populations
are
$\left< \log \Rint \right>= 0.0$
for \ERGs, and
$\left< \log \Rint \right>= -0.5$ 
for \NRGs.
For half of the \ERGs, $\Rint >1$, that is, the jet kinetic power is higher than the radiation from the accretion disk.
The physical properties of higher $\R$ sources 
are more clearly illustrated in Figure~\ref{fig:R_vs_sBHAR},
 which shows the distribution of ERGs and other WERGS sources in the $\Rint$--sBHAR plane.
The yellow shaded area is the region corresponding to $\lambdaedd=1$ assuming Equation~(\ref{eq:sBHAR_KH13})
for the range of $8<\log (\mstar/\msun)<12$.
The other WERGS sources are shown with a color gradation in $\log \R$,
with higher $\R$ lying toward the top right of the plot.
This means that higher $\R$ sources tend to have both higher $\lambdaedd$ and higher $\Rint$. 
It is well known that radio galaxies are more radio-loud as the Eddington-scaled
accretion rate decreases, producing the sequence from the top left region toward the bottom right one in Figure~\ref{fig:R_vs_sBHAR} \citep{ho02,mer07,pan07,sik07,sik13,ho08},
and overall \NRGs\ also follow this trend.
However, \ERGs\ are located in
the top right part of the plane, 
suggesting that 
\ERGs\ contain rapidly growing BH in the center, and at the same time harbor 
powerful, compact radio-jets.

These radio galaxies are different from the conventional radio galaxies in the local universe \citep{sey07}.
\cite{del18} recently found that high-$z$ ($z>1.5$) radio galaxies discovered
in the VLA/COSMOS survey tend to have
high Eddington ratios with $\log \lambdaedd > -2$ and that they are also starforming or starburst galaxies.
This is consistent with the properties of our overall WERGS sample
\citep[see also ][]{tob19a}, and thus \ERGs\ occupy the extreme end of highly accreting BHs in radio galaxies.

 \subsection{$\etaj$ and the Origin of Radio Emission in ERGs}
 \label{sec:figure7}
 
 Figure~\ref{fig:R_vs_sBHAR} indicates that \ERGs\ are a very interesting population,
 showing both radiative (high $\lambdaedd$) and jet-efficient (high $\Rint$) emission, 
which were missed in previous surveys.
 To investigate the jet properties in more detail, we here discuss the jet efficiency of \ERGs\ and compare the obtained values with other radio sources.
 The jet production efficiency is defined as
 \begin{equation}
 \etaj = \ljet / \mdotbh c^2 = \etar \ljet / \lbol 
 \end{equation}
 where $\etar$ is the radiation efficiency of an AGN accretion disk,
$c$ is the speed of light, and
$\mdotbh$ is the mass accretion rate onto the SMBH through the disk.
In the following, we adopt a 
canonical value of $\etar=0.1$
based on the Soltan-Paczynski argument \citep{sol82}.

The distribution of $\etaj$ is shown in the right panel of Figure~\ref{fig:hist_Rint}.
The median values are $\left< \log \etaj \right> = -1.0 \pm 0.3$ for \ERGs\ and
 $\left< \log \etaj \right> = -1.5 \pm 0.4$ for \NRGs.
 These values are slightly higher or consistent with the nearby radio AGN population ($\log \etaj \sim -1.5$) whose host galaxies are massive
\citep[$\mstar > 10^{11} \msun$;][]{nem15}.
 However, the local radio AGN population generally has
 much smaller Eddington ratios of $\log \lambdaedd < -2$.
 This population would be located in the top left region in Figure~\ref{fig:R_vs_sBHAR}, so they are clearly different from \ERGs.
The $\etaj$ of \ERGs\ is higher by a factor of $\sim10$ compared to the  SDSS radio quasar population, whose radio jet efficiency lies at $\left< \log \etaj \right> = -2.0 \pm 0.5$ \citep[e.g.,][]{van13b,ino17}.
Considering that the SDSS radio quasars on average show $\log \lambdaedd \sim -1$, those radio quasars would
be located at the bottom-center to bottom-right of Figure~\ref{fig:R_vs_sBHAR}, which is again different to the distribution of \ERGs.
Therefore, the disk-jet connection
 may be fundamentally different from the standard disk model, which conventionally describes the quasar/AGN accretion disk emission.

 One possible origin of this high production efficiency of radiation and jets is that the accretion disk of \ERGs\ is actually in the ``radiatively inefficient'' state, but the physical origin of radiative inefficiency is different from the disks in the local radio galaxies. For example, the \ERGs\ may be undergoing more rapid mass accretion \citep[so called slim disk model;][]{abr88}. 
 Recent radiation hydrodynamical simulations suggest that when the mass accretion rate significantly exceeds the Eddington rate,
radiation is effectively trapped within the accreting matter and advected to the central BH before escaping by radiative diffusion. 
As a result, the emergent radiation luminosity is saturated at the order of $\ledd$ (i.e., $\etar \lesssim 0.1$ at $\gg \dot{M}_\mathrm{Edd}$) 
and the accretion flow turns into a radiatively inefficient state even with a super-Eddington accretion rate \citep[e.g.,][]{ohs05,ohs09,mck14,ina16b,tak20}.
 This phase is considered to be a key process of the BH seed growth in the high-$z$ ($z>6$) universe to describe the already known massive high-$z$ SMBHs \citep[e.g.,][]{mor11,wu15,ban18,ono19}.

In the situation that the accretion rate is well above the Eddington rate, 
a relativistic jet can be launched by the Blandford-Znajek mechanism
if large-scale magnetic fields exist in the innermost region \citep{bla77,tch11}.
The jet behavior of these extreme cases
  has been investigated by several authors \citep{sad14,sad15}
  using a 3D general relativistic radiation magneto-hydrodynamical simulation (GRRMHD) code.
 They find that,
  in these (super-)Eddington accretion phase, BHs emit a significant amount of radiation and jet energy, and both the radiation and jet power approach around the Eddington limit, which appears to be the case for \ERGs\ and the sources in the top right region in Figure~\ref{fig:R_vs_sBHAR} \citep[see also][]{bla19}.
 Thus, the study of \ERGs\ at $z\sim1$ will be complementary to studies utilizing the next generation instruments capable of observing seed BHs in the early universe at $z \gtrsim 7$ \citep{hai20} in the investigations of the growth of seed BHs.

\begin{figure}
\begin{center}
\includegraphics[width=0.47\textwidth]{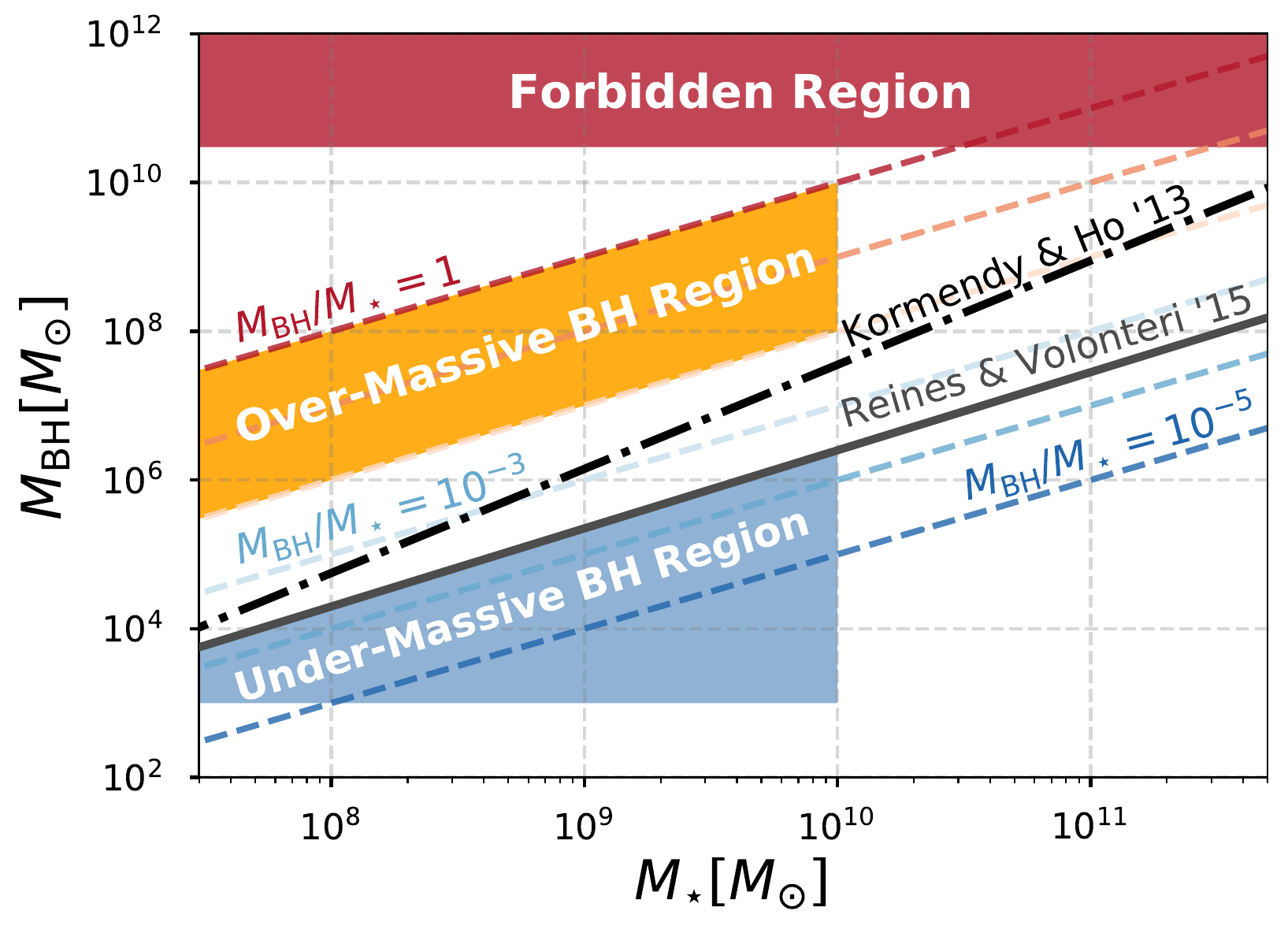}\\
\caption{
A schematic view of the relationship between $\mbh$ and $\mstar$.
The local relations by \cite{kor13} and \cite{rei15} are shown as dot-dashed (black; the fiducial relation in this study) and solid-line (gray), respectively.
The orange shaded region represents the range of over-massive BH, which is at least one order of magnitude higher than the local relations
(Case 1 in Section~\ref{sec:alternative}),
and the region is delineated by $10^8 < \mstar/\msun < 10^{10}$ and $10^{-2}<\mbh/\mstar < 1$. 
The blue shaded region represents the range of under-massive BH, in which  host galaxies form first, and BH grow later (Case 2 in Section~\ref{sec:alternative}). 
This region is delineated by $10^8 < \mstar/\msun < 10^{10}$ and $10^{-2}<\mbh/\mstar$ and smaller than the local relation by \cite{rei15}.
The red shaded region represents the forbidden region where the BH cannot exceed $\mbh \lesssim 3\times10^{10} \msun$ \citep{ina16,ich17b}.
}\label{fig:MBH_vs_Mstar}
\end{center}
\end{figure}

\subsection{Alternative Scenarios: If ERGs Do Not Follow Local Scaling Relations between $\mbh$ and $\mstar$}\label{sec:alternative}

The discussion related to the Eddington ratio depends on the $\mbh$ estimation, which in turn relies on the
local scaling relation of \cite{kor13}. 
However, it is not obvious that this relation also holds in the low-mass end in which many \ERGs\ probably reside.
Although the current data cannot give any constraints on this issue and a more detailed study is beyond the
scope of this paper, we discuss two alternative scenarios, in which 
our galaxies are above (case one) or below (case two) the correlation between $\mbh$ and $\mstar$ as shown in Figure~\ref{fig:MBH_vs_Mstar}.

Case one reflects the possibility that \ERGs\ might host over-massive BHs
 compared to the relation by \cite{kor13}, with the BH-to-stellar-mass ratio $\fbstar \equiv \mbh/\mstar \gtrsim  10^{-2}$
 (see the orange shaded area in Figure~\ref{fig:MBH_vs_Mstar}).
In this case, concerns surrounding super-Eddington accretion discussed in Section~\ref{sec:sBHAR} and \ref{sec:highR} are alleviated: 
the median Eddington ratio for \ERGs\ can be parametrized by $\langle \lambdaedd \rangle \simeq
0.17~(\fbstar/0.01)^{-1}$.

In addition, the possible future growth path in the plane of $\mbh$ and $\mstar$ can be discussed by assuming the current BH/stellar mass assembly rate.
Since the radiation efficiency is $ \etar = 0.1$ in the sub-Eddington regime, 
the BH accretion rate is estimated as $\mdotbh = \lbol/(\etar c^2) \simeq 1.7 \times 10^{-3} \fbstar^{-1} \dot{M}_\mathrm{Edd}$,
where the median Eddington ratio above is substituted
(note that this equation is valid at $\fbstar \gtrsim 1.7 \times 10^{-3}$).
Comparing the BH accretion rate to the sSFR of \ERGs\ (sSFR$=\mathrm{SFR}/\mstar \simeq 10^{-8}~\mathrm{yr}^{-1}$; see Figure~\ref{fig:SFRMstar}),
we obtain $\dot{M}_\mathrm{BH}/\mathrm{SFR} \simeq 4 \times 10^{-3}$, which is independent of the choice of $\fbstar$.
Therefore, the BH-galaxy mass ratio for \ERGs\ is expected to approach a canonical value observed in the local universe \citep[e.g., $\fbstar \sim 3\times 10^{-3}$;][]{kor13}.
This evolutional path might be analogous to that of known high-$z$ luminous quasars whose SMBHs are likely over-massive compared to the local relation
with $10^{-2} \lesssim \fbstar \lesssim 2\times 10^{-1}$ \citep[e.g.,][]{wan13,wan16,tra15,dec18}.

Case two considers the possibility that \ERGs\ might host very under-massive BHs with $\fbstar \lesssim 10^{-3}$--$10^{-4}$ as shown in the blue shaded region in Figure~\ref{fig:MBH_vs_Mstar}.
For instance, the overall normalization for late-type galaxies may be considerably lower than for the local relation of early-type galaxies
\citep[e.g.,][]{rei15,las16,gre20}.
This ``galaxy grows first, BH come later'' phase may occur because gas accretion onto the BHs is quite inefficient 
due to efficient stellar feedback, e.g., star formation-driven outflows, and the shallow halo potential until the stellar-mass reaches a critical value of $\mstar \sim 10^{10.5} \msun$ \citep[e.g.,][]{bow17,hab17}. 
However, once the galaxy reaches the critical mass, above which star formation-driven outflows no longer prevent the gas accretion to the galactic nuclei, they abruptly switch to a rapid growth phase in the absence of feedback, and finally merge into the local relation. 
\ERGs\ with $\mstar \gtrsim 10^{10.5} \msun$ might be in such a rapidly growing phase.

The scenario above looks promising but it is worth
discussing the critical mass value because the stellar mass of most \ERGs\ actually lies below the expected critical mass. 
Beside, Figure~\ref{fig:sBHARvsMstar} shows that the median sBHAR (and therefore $\lambdaedd$) tends to be higher for less massive \ERGs\ with
the median Eddington ratio of $\langle \lambdaedd \rangle \simeq 1.7~(\fbstar/10^{-3})^{-1}$, suggesting that the extreme BH growth is possible (at least temporally) even in lower mass galaxies.
This discrepancy may reflect that there are missing understanding on the physics of AGN feedback and gas feeding in low mass galaxies.

Another interesting point in the second scenario is that the Eddington ratio for several \ERGs\ becomes as high as $\lambdaedd \sim 10^2$.
Recent numerical simulations suggest that for a highly magnetized accretion disk around a rapidly spinning BH, the disk transits into 
a magnetically arrested disk (MAD) state and produces high radiative luminosity \citep[e.g.,][]{mck14,sad15}.
Through the emission process, the level of high luminosity could be achieved only when the BH mass accretion exceeds 
$\sim 10^3~\dot{M}_\mathrm{Edd}$ (\citealt{ina16b}; see also Figure 5 in \citealt{ina20}).
Therefore, if this scenario is true, future follow-up observations of those \ERGs\ with such high Eddington ratios
would reveal the properties of possibly hyper-Eddington accreting BHs \citep{tak20}.

\subsection{Prospects for AGN Feedback in Low-Mass Galaxies through \ERGs}\label{sec:lowmassRG}

In recent years, the theoretical study of AGN feedback has been extended to low-mass galaxies \citep{sil17}. 
This was motivated mainly by the increasing number of intriguing observational results showing signs of AGN feedback in low-mass galaxies \citep{nyl17,pen18,kav19,dic19,mez19}.
Our results show that the jet power of $\ljet > 10^{44}$~erg~s$^{-1}$ in \ERGs\ is equivalent to that produced from radio quasars and local radio AGN residing in massive galaxies. 
This suggests the presence of a strong jet that can disperse or even expel the interstellar medium \citep[e.g.,][]{mor05, hol08, nes17} or produce
cavities in the host galaxies, which are often observed in the local massive counterparts \citep[e.g.,][]{raf06,mcn07,bir08,bla19}.
On the other hand, low stellar-mass \ERGs\ are still in a starburst phase, indicative of a substantial gas reservoir,
and lacking clear signatures of negative AGN feedback\footnote{Numerical simulations suggest that mechanical feedbacks associated with jets would impact upon their host galaxies with a significant delay time, which might be marginally longer than the typical AGN lifetime \citep{cos18}}.

To investigate whether such a negative feedback from the jet is effective in the host galaxies, it is necessary to obtain high spatial resolution radio images that resolve the host galaxy 
down to scales of $<10$~kpc \citep{rei20}. Our sample of compact radio sources from VLA/FIRST with a spatial resolution of $\sim5$~arcsec only provide poor upper-bounds of the radio jet size of $<40$~kpc at $z\sim1$. Future higher quality observations with sub-arcsecond resolution will provide more insight into the effect of the jet on the host galaxy.

Given that the BH mass of \ERGs\ might reach the massive BH range below $\mbh \sim 10^6 \msun$,
studying the properties of \ERGs\ gives a window into understanding seed BH growth in the high-$z$ universe.
Some observational studies recently suggested that AGN feedback might also have \textit{positive} effect on the host galaxies,
and intensive in situ star formation is ignited in the outflowing gas launched by the AGN \citep{mai17}.
Recent numerical simulations of jet-driven feedback also predict star-formation triggered on galaxy scales by the overpressure of the jet during the first few Myrs of strong interaction between the jet and the interstellar medium \citep{wag11, wag16, gai12, muk18}.
If the radio jets in low-mass galaxies have a positive impact on both star formation and BH growth,
then radio-selected AGN in low-mass galaxies are an intriguing population for understanding the BH and stellar mass assembly in the environment of shallow potential low-mass galaxies.

\subsection{Wide Area Radio and Optical Surveys; Potential for Discovering Previously Missed Populations}

The small number density of \ERGs\ ($\sim 1$ source deg$^{-2}$) and their optical faintness of $i_\mathrm{AB} \sim 25$ suggest that \ERGs\ 
were easily missed in previous optical surveys of radio sources. Our work shows that \ERGs\ can be discovered with wide ($>100$~deg$^2$) and deep ($i_\mathrm{AB} \lesssim 26$) optical follow-up observations of radio sources.
The number of known \ERGs\ will increase in near future,
once the Subaru/SSP survey is complete, covering an area of $\sim10^3$~deg$^2$. The upcoming Subaru/Prime Focus Spectrograph \citep[PSF; ][]{tak14} will conduct extensive optical and near-IR spectroscopic follow-up observations in the footprint of the HSC-SSP field, providing spec-$z$ information of \ERGs\ with $i_\mathrm{AB} \lesssim 24$.
The forthcoming LSST survey \citep{ive19} will cover half of the sky, and it will also increase the number of sources by another order of magnitude, resulting in $\sim 10^4$ sources.
A significant fraction of these \ERGs\ will also be expected to reside in low-mass galaxies.
Therefore, the combination of wide area (either shallow or deep) radio and deep optical imaging and spectroscopic surveys will give us 
a statistically large number of massive BH candidates in the coming years.

Massive BH searches have been conducted 
with multi-wavelength data;
optical spectroscopy using broad H$\alpha$ emission \citep{gre04,gre07,xia11} or narrow emission line ratios \citep[e.g.,][]{bar08,rei13}, 
mid-IR ($W1-W2$) colors in \textit{WISE} \citep{sat14,sar15,sec15,kav19},
X-ray observations
\citep[e.g.,][]{sch13c,bal15,mez16,che17,kaw19},
as well as through studies of the AGN flux variability
based on the properties that lower luminosity AGN have stronger variability amplitude \citep{mor16,bal18,bal19,elm20,kim20}.
On the other hand, radio surveys have rarely been used for searching massive BHs in low-mass galaxies. 

Recently, \cite{rei20} employed the radio-survey approach to search for massive BHs by using VLA/FIRST and conducting VLA high spatial resolution follow-up observations
 of local dwarf galaxies at $z<0.1$, and found more than 10 candidates. One remarkable result was that most radio cores were not located
 at the center of the host, but off-nucleus, 
a possible signature of a previous merger.
If the detection of an off-nucleus core is indeed the consequence of a galaxy merger, 
it indicates that the orbital decay of a black hole through dynamical friction is inefficient in this BH mass range, resulting in a population of BHs that fail to reach the galactic center within a Hubble time, since major mergers in dwarf galaxies becomes rare at $z<3$ \citep{fri18}.
Currently there are many theoretical implication regarding wandering BHs \citep{com14,ble16,tre18,der20,guo20,ric21a,ric21b}.
High spatial resolution radio imaging can pin down the locations of such wandering BHs, and large-scale statistical studies may provide insights into the frequency and environment of wandering BHs through the combination of LSST and VLA/FIRST or the ongoing VLASS \citep{lac20}, which will achieve a sensitivity down to $0.1$~mJy at 2--4~GHz, giving an order of magnitude deeper radio observations than VLA/FIRST.

\subsection{Do Blazars Contaminate ERGs?}

It could be argued that blazars might contaminate the sample of \ERGs, and that therefore the optical emission is dominated by the jet component.
Based on the known blazar sequence
in the radio and optical bands \citep{fos98,don01,ino09,ghi17}, 
the observed radio-loudness of blazars is in the range of $\log \mathcal{R}_\mathrm{obs} \sim 1.5-3$. This range is obtained 
by using the luminosity range of $\lfirst=10^{41-43}$~erg~s$^{-1}$ and
shifting redshift to the range of $z\sim0$--$5$. 
Therefore, blazars cannot reproduce extreme radio-loudness reaching $\log \R > 4$. 
Blazar contamination is unlikely
also from the point of the observed optical magnitude.
Considering that all ERGs are VLA/FIRST selected with a flux limit of $>1$~mJy, the expected observed magnitude of blazars based on the blazar sequence is high with a peak around $i_\mathrm{AB}=21$,
and all cases are at $i_\mathrm{AB} < 23.5$
even when changing the luminosity range of the blazar SEDs and also shifting the redshift range to $z=0$--$5$.
Based on these arguments, the contamination of blazars to \ERGs\ is unlikely.

\section{Conclusion}\label{sec:conclusion}

The long-missing optical counterparts of radio-bright ($f_\nu > 1$~mJy at 1.4~GHz) VLA/FIRST population
have been constructed by \cite{yam18} by utilizing the wide ($>100$~deg$^2$) and deep ($i_\mathrm{AB} < 26$) Subaru/HSC SSP survey
(named WERGS). The WERGS catalog contains a unique radio galaxy population with $\log \R>4$, or extremely radio-loud galaxies (\ERGs).
The number density of \ERGs\ indicates that they are a very rare population ($\sim1$ source deg$^{-2}$),
and therefore it is not surprising that \ERGs\ were missed in previous optical counterpart surveys of the radio sources.
Combining multi-wavelength data in optical, IR, and radio bands \citep{tob19a},
we have compiled a sample of \Nergs~\ERGs\ and \Nwergsother~\NRGs\ with radio-loudness is $1< \log \R<4$.

We have investigated the properties of these unique \ERGs\ and found the following key results:

\begin{enumerate}

\item Although the estimated SFRs
are sensitive to assumed SFHs, and they
might have large uncertainties with $\sim0.5$~dex, all \ERGs\ are likely in the star-forming or starburst phase reaching specific star-formation rate of $\mathrm{sSFR}\sim 10^{-8}$~yr$^{-1}$, suggesting that \ERGs\ might be in a rapid stellar-mass assembly phase with mass doubling times of $\sim100$~Myr. Besides, their stellar mass is relatively small, including the low-mass galaxies with $\mstar <10^{10} \msun$.
\item IR detected \ERGs\ are in a rapid BH accretion phase with high specific black hole accretion rate (sBHAR) with the expected Eddington ratio $\left< \log \lambdaedd \right> \approx -0.4$, and some \ERGs\ may be experiencing a super-Eddington phase.
\item Sources with higher $\R$ tend to have both higher sBHAR (and thus higher $\lambdaedd$) and higher intrinsic radio-loudness $\Rint$. This paints a different picture of radio galaxies compared to conventionally known local radio galaxies with low $\lambdaedd$. \ERGs\ represent a population of unique radio galaxies characterized by both high $\lambdaedd$ and high radio power.
\item The jet efficiencies of \ERGs\ are typically $\etaj \sim 0.1$, which is similar to local radio AGN residing in massive galaxies, 
whereas it is an order of magnitude higher than the values estimated for SDSS radio-loud quasars. This indicates that, although
the accretion disk is reaching super-Eddington accretion, it is in 
a radiatively inefficient phase (possibly in a slim disk configuration) due to rapid mass accretion onto the central BH.
\end{enumerate}

\acknowledgments

We thank the anonymous referee for helpful suggestions that strengthened the paper.
We thank Yoshiyuki Inoue for providing us
the blazar sequence SED templates.
This work is supported by Program for Establishing a Consortium
for the Development of Human Resources in Science
and Technology, Japan Science and Technology Agency (JST) and is partially supported by Japan Society for the Promotion of Science (JSPS) KAKENHI (18K13584 and 20H01939; K.~Ichikawa, 18J01050 and 19K14759; Y.~Toba, 16H03958, 17H01114, 19H00697; T.~Nagao, 17J09016: T.~Kawamuro, 19K03862: A.~Y.~Wagner).
M.~Charisi acknowledges support from
the National Science Foundation (NSF) NANOGrav Physics Frontier Center, award number 1430284.

The Hyper Suprime-Cam (HSC) collaboration includes the astronomical communities of Japan and Taiwan, and Princeton University.  The HSC instrumentation and software were developed by the National Astronomical Observatory of Japan (NAOJ), the Kavli Institute for the Physics and Mathematics of the Universe (Kavli IPMU), the University of Tokyo, the High Energy Accelerator Research Organization (KEK), the Academia Sinica Institute for Astronomy and Astrophysics in Taiwan (ASIAA), and Princeton University.  Funding was contributed by the FIRST program from the Japanese Cabinet Office, the Ministry of Education, Culture, Sports, Science and Technology (MEXT), the Japan Society for the Promotion of Science (JSPS), Japan Science and Technology Agency  (JST), the Toray Science  Foundation, NAOJ, Kavli IPMU, KEK, ASIAA, and Princeton University.

This paper makes use of software developed for the Large Synoptic Survey Telescope. We thank the LSST Project for making their code available as free software at  http://dm.lsst.org

This paper is based on data collected at the Subaru Telescope and retrieved from the HSC data archive system, which is operated by Subaru Telescope and Astronomy Data Center (ADC) at NAOJ. Data analysis was in part carried out with the cooperation of Center for Computational Astrophysics (CfCA), NAOJ.

The Pan-STARRS1 Surveys (PS1) and the PS1 public science archive have been made possible through contributions by the Institute for Astronomy, the University of Hawaii, the Pan-STARRS Project Office, the Max Planck Society and its participating institutes, the Max Planck Institute for Astronomy, Heidelberg, and the Max Planck Institute for Extraterrestrial Physics, Garching, The Johns Hopkins University, Durham University, the University of Edinburgh, the Queen’s University Belfast, the Harvard-Smithsonian Center for Astrophysics, the Las Cumbres Observatory Global Telescope Network Incorporated, the National Central University of Taiwan, the Space Telescope Science Institute, the National Aeronautics and Space Administration under grant No. NNX08AR22G issued through the Planetary Science Division of the NASA Science Mission Directorate, the National Science Foundation grant No. AST-1238877, the University of Maryland, Eotvos Lorand University (ELTE), the Los Alamos National Laboratory, and the Gordon and Betty Moore Foundation.

%

\vspace{5mm}
\facilities{VLA, \textit{Herschel}, \textit{WISE}, Subaru/HSC}


\software{astropy \citep{ast13}, Matplotlib \citep{hun07}, Pandas \citep{mck10}
          }



\appendix


\section{How Much Do Wrong Photo-$z$
Affect Our Results?}\label{sec:appendix}

\begin{figure*}
\begin{center}
\includegraphics[width=0.5\textwidth]{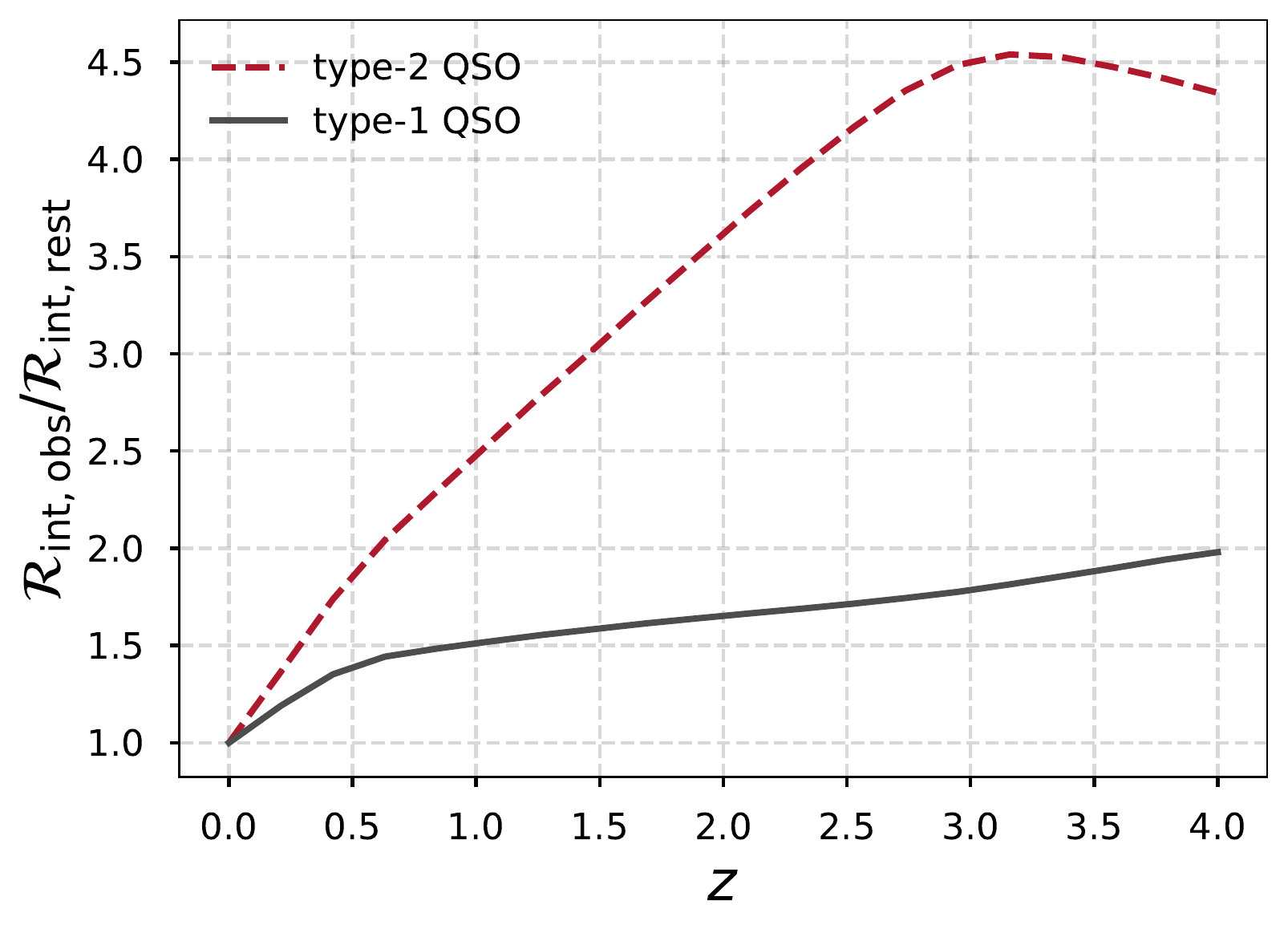}~
\includegraphics[width=0.5\textwidth]{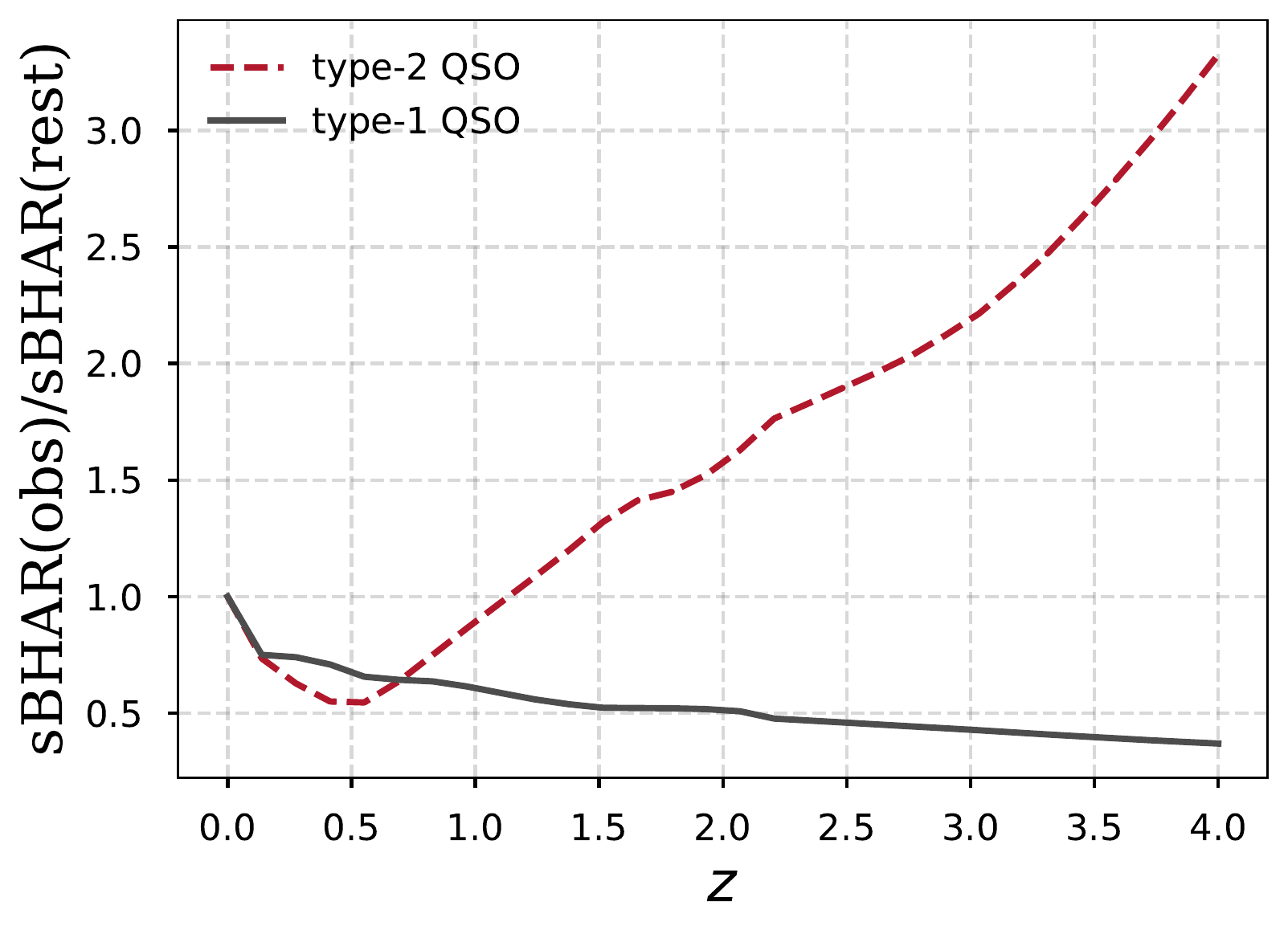}
\caption{
The $k$-correction for different physical quantities as a function of redshift. 
The SED template obtained by \cite{hic17} (type-1: black solid line, and type-2: dashed red line) is used for the $k$-correction estimation.
(Left) $\mathcal{R}_\mathrm{int,obs}/\mathcal{R}_\mathrm{int,rest} = (f_\mathrm{\nu, 1.5 GHz, obs}/f_\mathrm{\nu, 12 \mu m, obs})/(f_\mathrm{\nu, 1.5 GHz, rest}/f_\mathrm{\nu, 12 \mu m, rest})$.
(Right) sBHAR(obs)/sBHAR(rest)$=(f_\mathrm{\nu, 12 \mu m, obs}/f_{\nu, i~\mathrm{band, obs}}) / (f_\mathrm{\nu, 12 \mu m, rest}/f_{\nu, i~\mathrm{band, rest}})$.
}\label{fig:obs_rest_vs_z}
\end{center}
\end{figure*}

Our results on \ERGs\ rely on photo-$z$ estimates, which affect physical values calculated using luminosity distance, e.g., stellar-mass and luminosities. 
Unfortunately, because most of our sources are optically faint, there are no spec-$z$ confirmed \ERGs\ in our sample.
A concern is how erroneous photo-$z$ estimates may affect our results.

If our photo-$z$ values are severely under-estimated, it is clear that the absolute physical values such as stellar-mass, with the dependence of luminosity distance of $D_L^2$, have been underestimated, and therefore the results on low-mass BH mass and low-mass galaxy arguments discussed in Section~\ref{sec:lowmassRG} are no longer valid.
One situation this might happen is that our photo-$z$ estimation at $z\sim0.3$--$0.5$ is not tracing the Balmer break at $\sim4000$~\AA, but actually the Lyman break at $z\sim3$--$4$ \citep[e.g.,][]{tan18}.

On the other hand, physical quantities obtained from the ratios of two physical quantities have weaker redshift dependences.
For both $\Rint$ and sBHAR, the luminosity distance dependence is canceled out.
The remaining $z$-dependent factor is the $k$-correction between the observed flux and the rest frame flux.
The physical quantities $\ljet$, $\liragn$ (and therefore $\lbol$), and $\mstar$, can be mainly traced by one-band flux (densities) of FIRST 1.5~GHz tracing radio Synchrotron emission ($f_\mathrm{\nu, 1.5GHz}$), \textit{WISE} MIR band(s)
tracing the peak of AGN dust emission ($f_{\nu,12\mu \mathrm{m}}$), and $i$-band tracing stellar emission ($f_{\nu,i~\mathrm{band}}$). 
Therefore, each physical quantity can be written as
$\Rint = \ljet/\lbol \propto f_\mathrm{\nu, 1.5GHz}/f_{\nu, 12\mu \mathrm{m}}$ and sBHAR$=\lbol / \mstar \propto f_{\nu,12\mu \mathrm{m}}/f_{\nu,i~\mathrm{band}}$. 
We checked and confirmed the proportionality of these ratios in our sample.

We then investigated the $k$-correction by seeeing how much these observed values change as a function of redshift.
In the radio-band, we assumed the same power-law $f_\nu \propto \nu^\alpha$ with $\alpha=-0.7$.
In the optical and IR band, we used the quasar SED templates obtained by \cite{hic17}, which contain two templates of type-1 and type-2 quasars, whose optical band is dominated by accretion disk and stellar-component, respectively. Since the SED extends only down to 2000~\AA,
we extrapolated the SED with $\alpha=-0.5$ for type-1 QSOs and $\alpha=-2.4$ for type-2 QSOs in order to smoothly connect the SED template \citep[eg., see also][for other SED templates]{ric06,pol07}. 
Given their high radio-loudness and as we discussed in Section~\ref{sec:analysis}, the SED of \ERGs\ is likely similar to type-2 QSOs, whose optical emission is dominated by the stellar-component.

Figure~\ref{fig:obs_rest_vs_z} shows the $k$-correction of each physical value as a function of redshift. The $k$-correction changes by a factor of
up to 4.5 for $\Rint$ and factor of $3$ for sBHAR even when the targets are shifted at $z\sim 4$. This suggests that an erroneous photo-$z$
does not strongly affect our overall results for $\Rint$ and sBHAR.

\section{Redshift dependence of our results}\label{sec:appendix_zdep}

\begin{figure*}
\begin{center}
\includegraphics[width=0.33\textwidth]{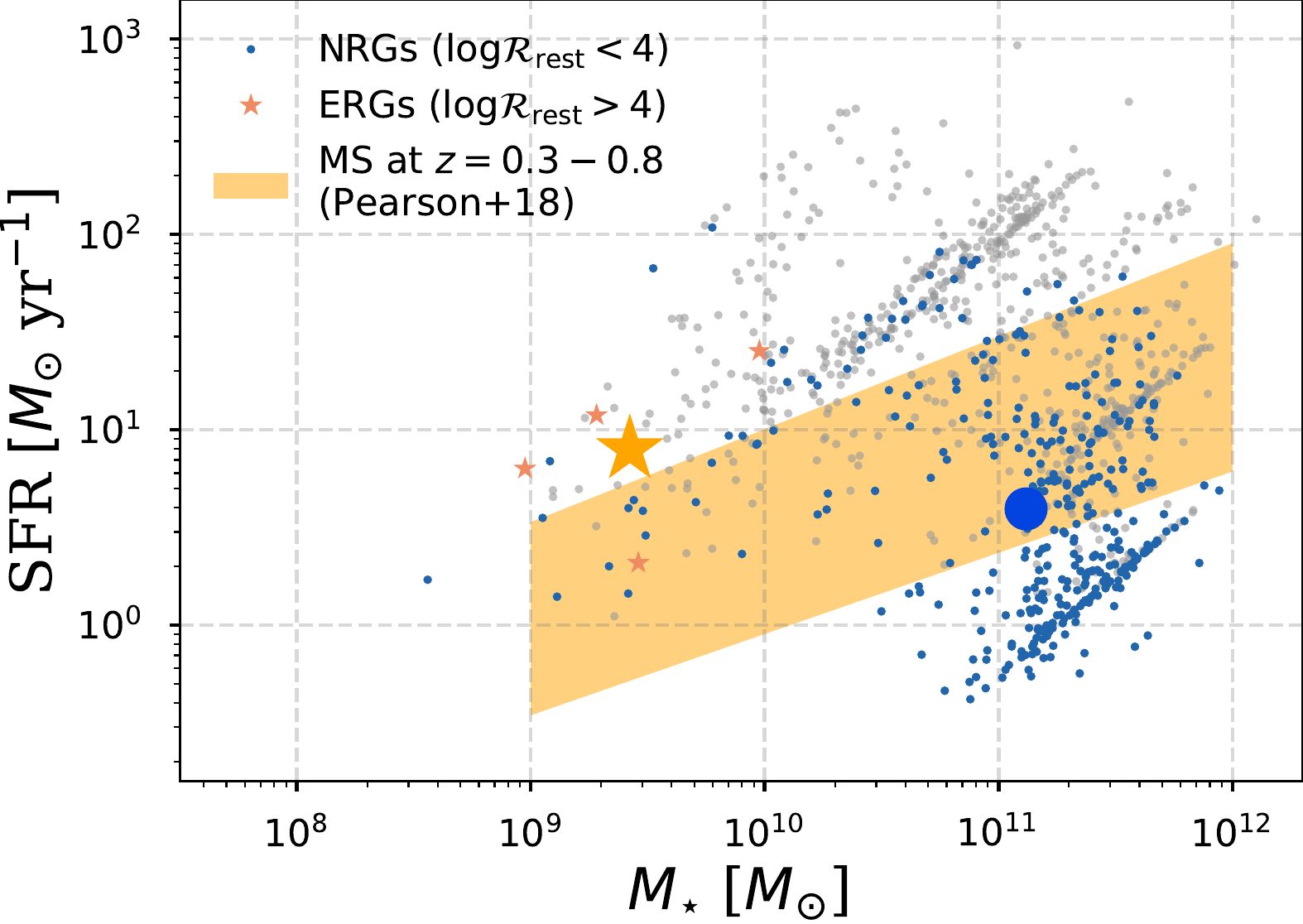}~
\includegraphics[width=0.33\textwidth]{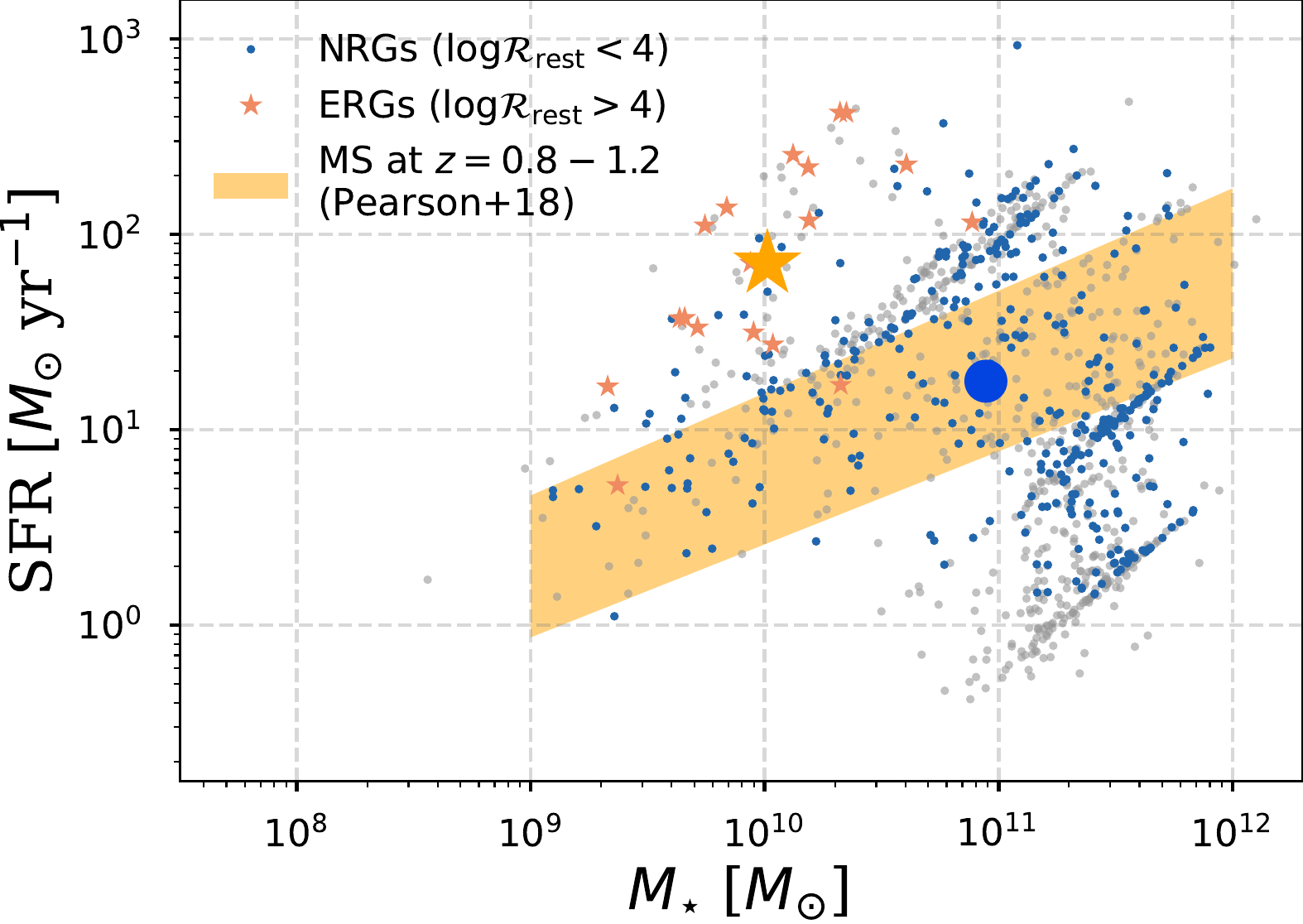}~
\includegraphics[width=0.33\textwidth]{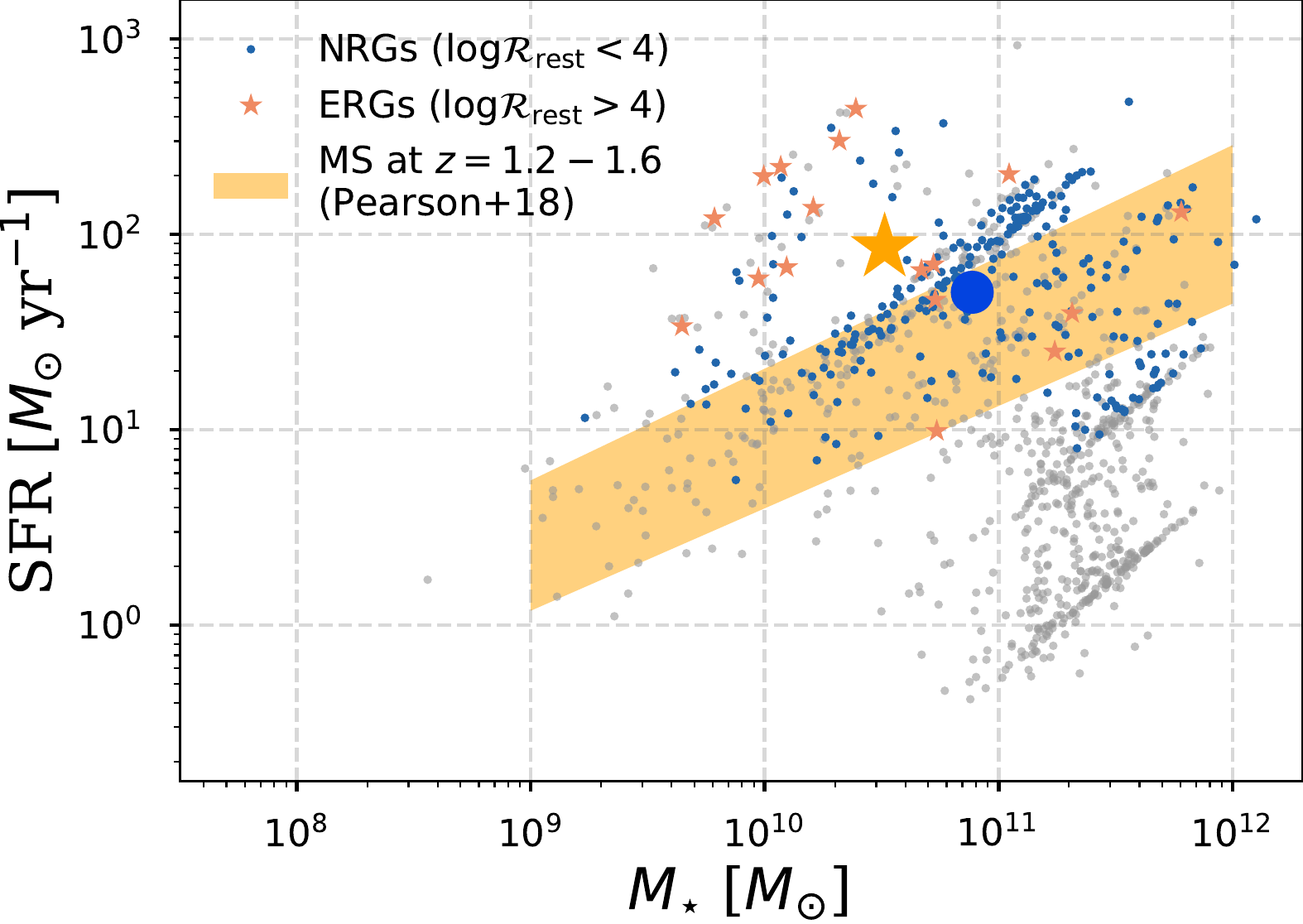}\\
\includegraphics[width=0.33\textwidth]{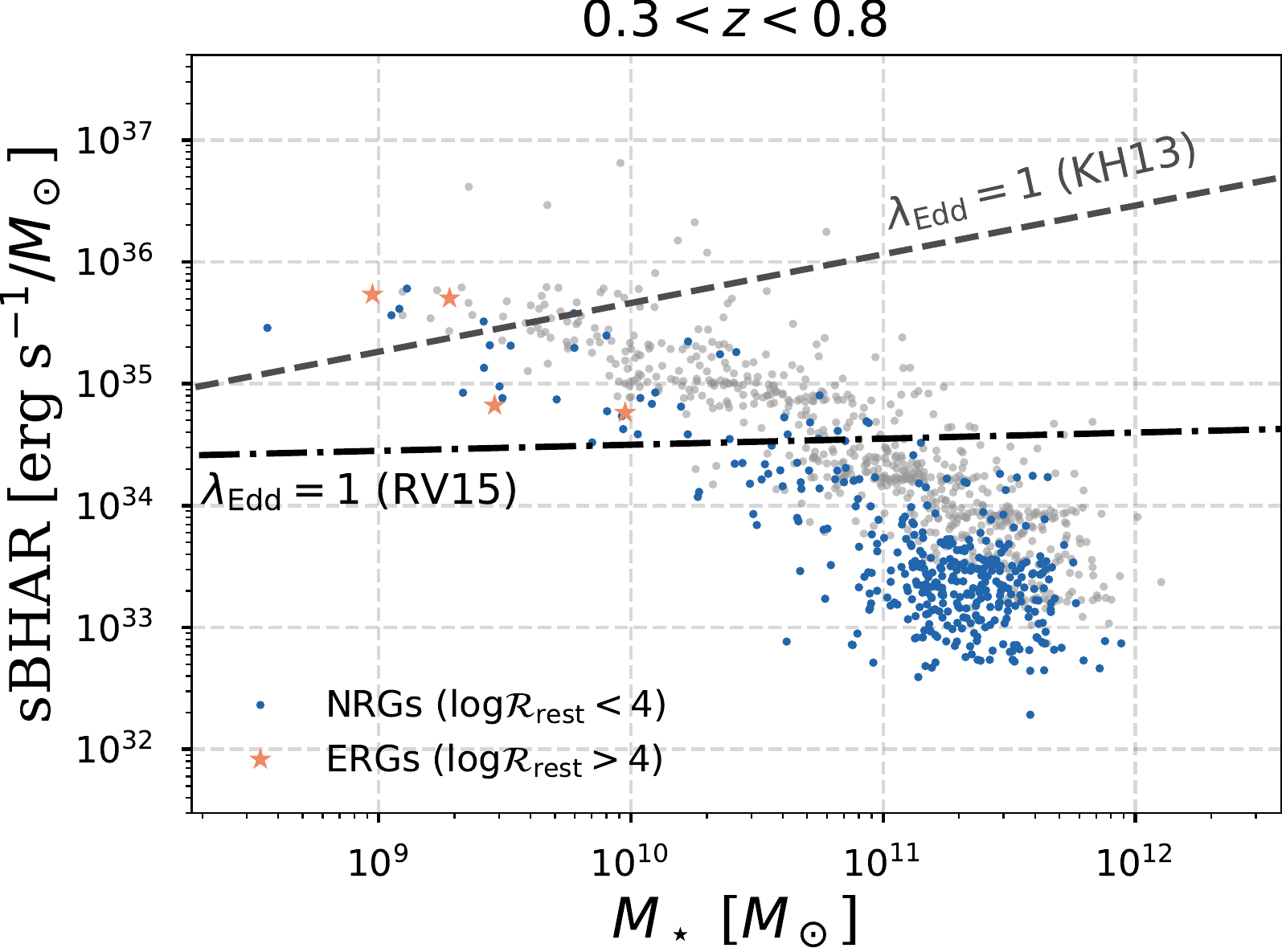}~
\includegraphics[width=0.33\textwidth]{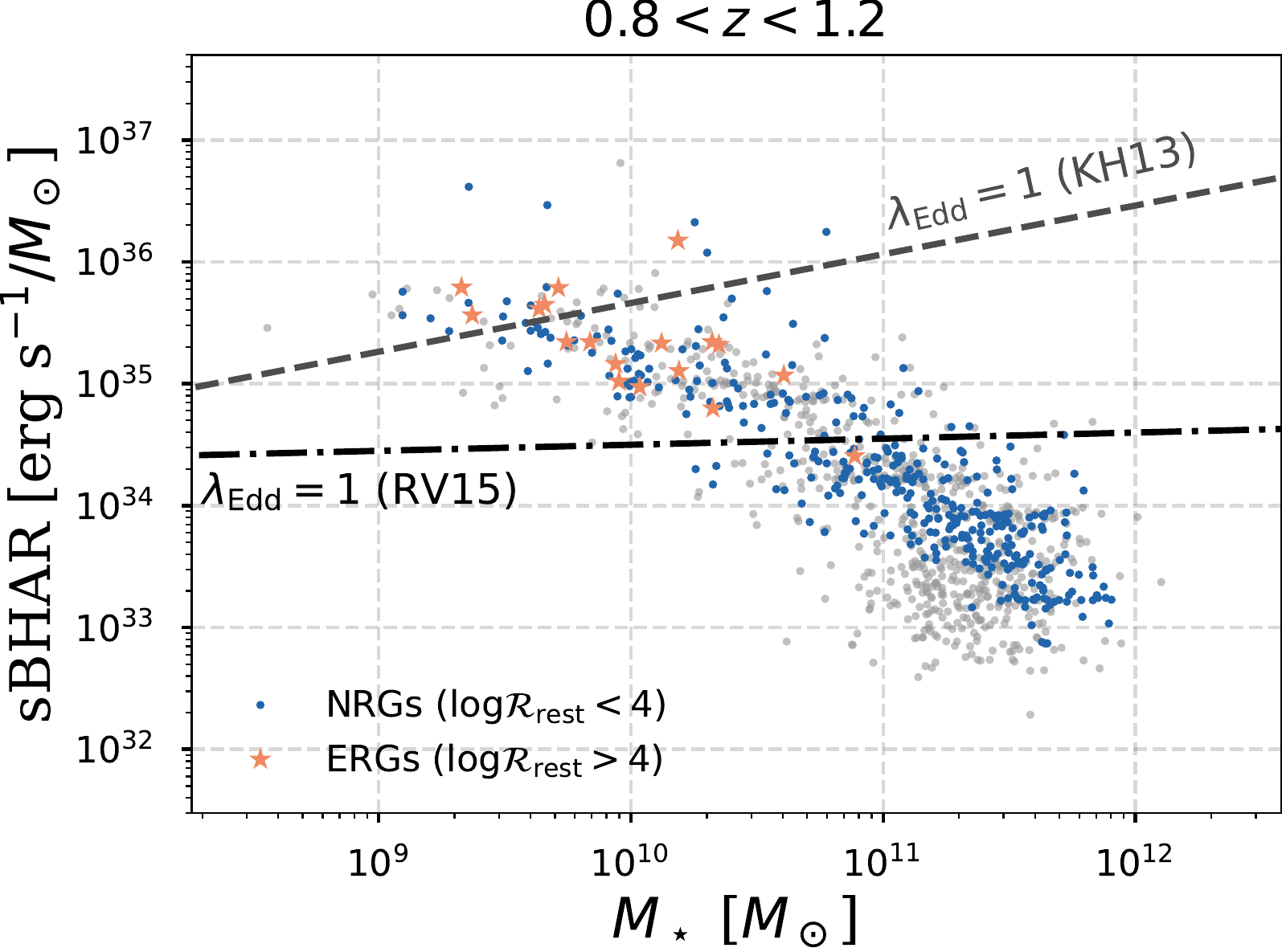}~
\includegraphics[width=0.33\textwidth]{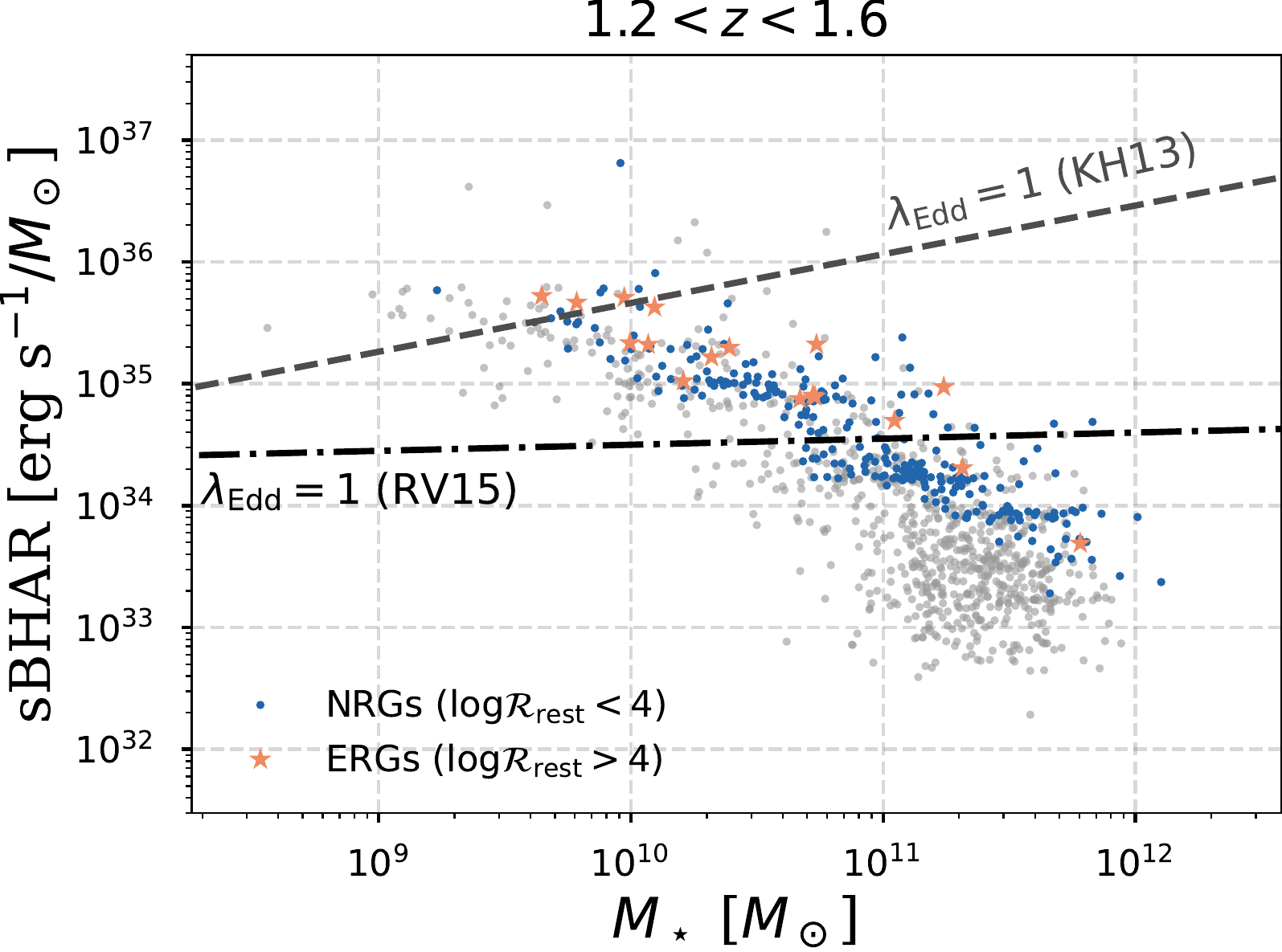}\\
\includegraphics[width=0.33\textwidth]{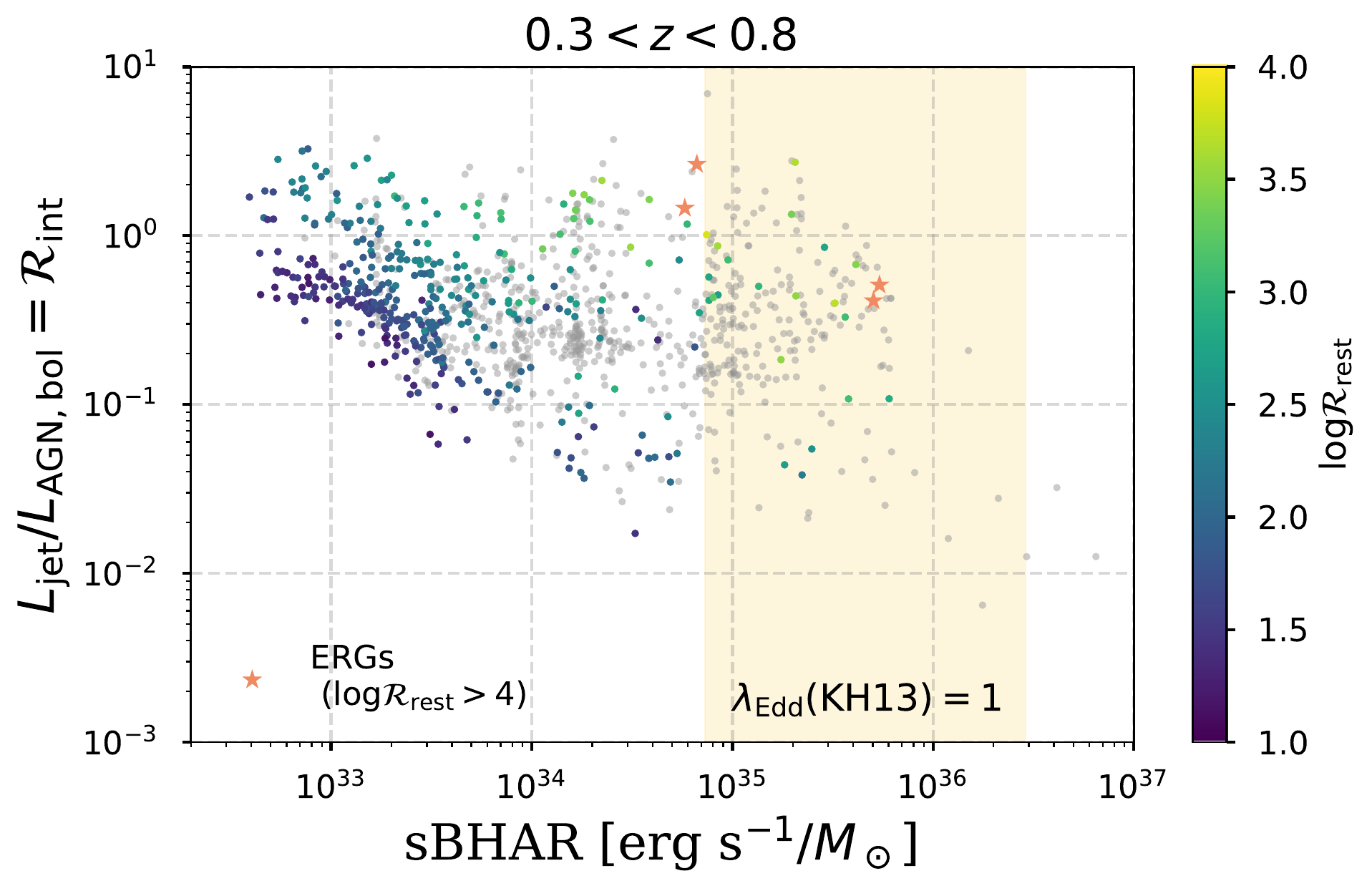}~
\includegraphics[width=0.33\textwidth]{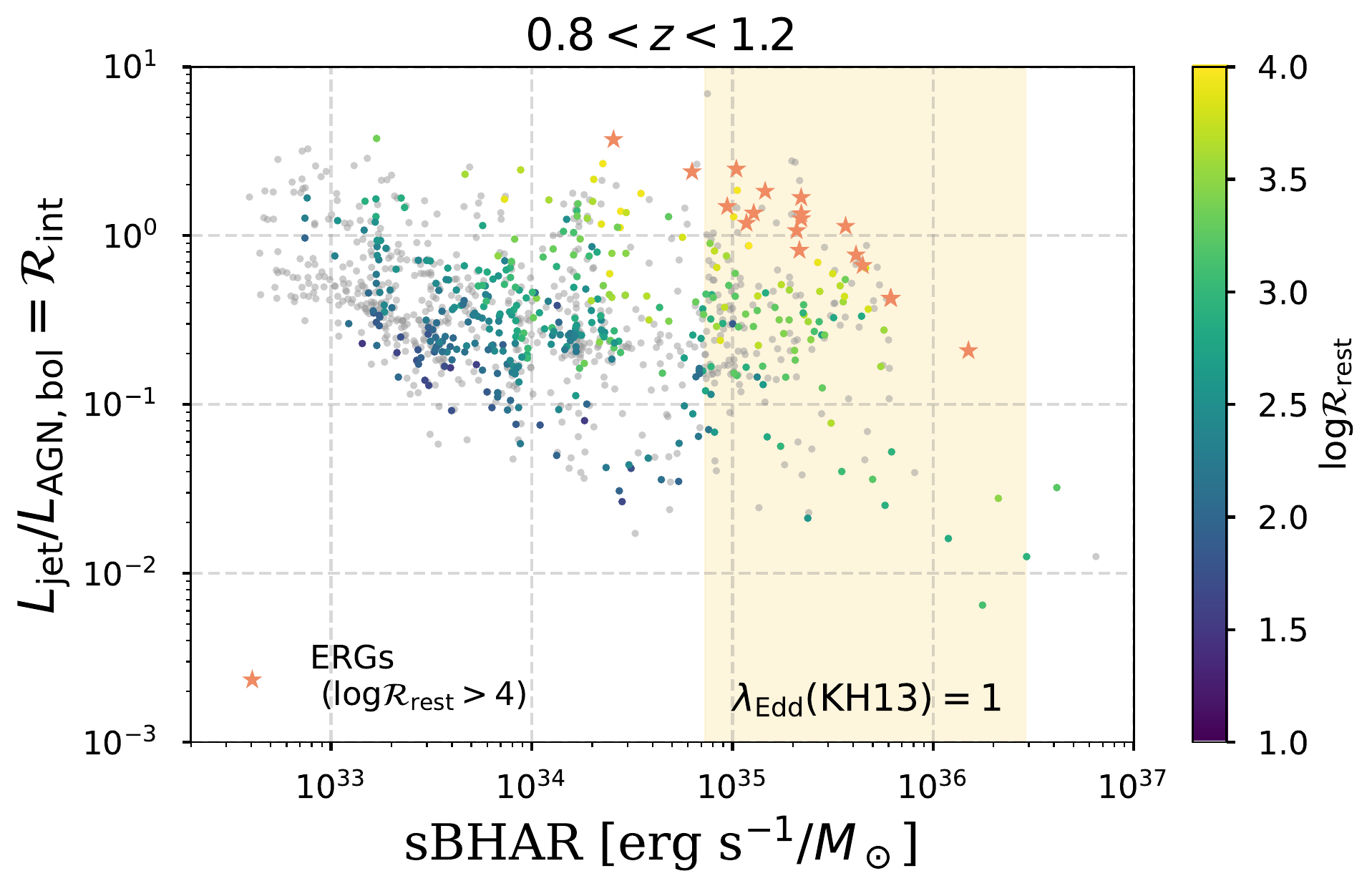}~
\includegraphics[width=0.33\textwidth]{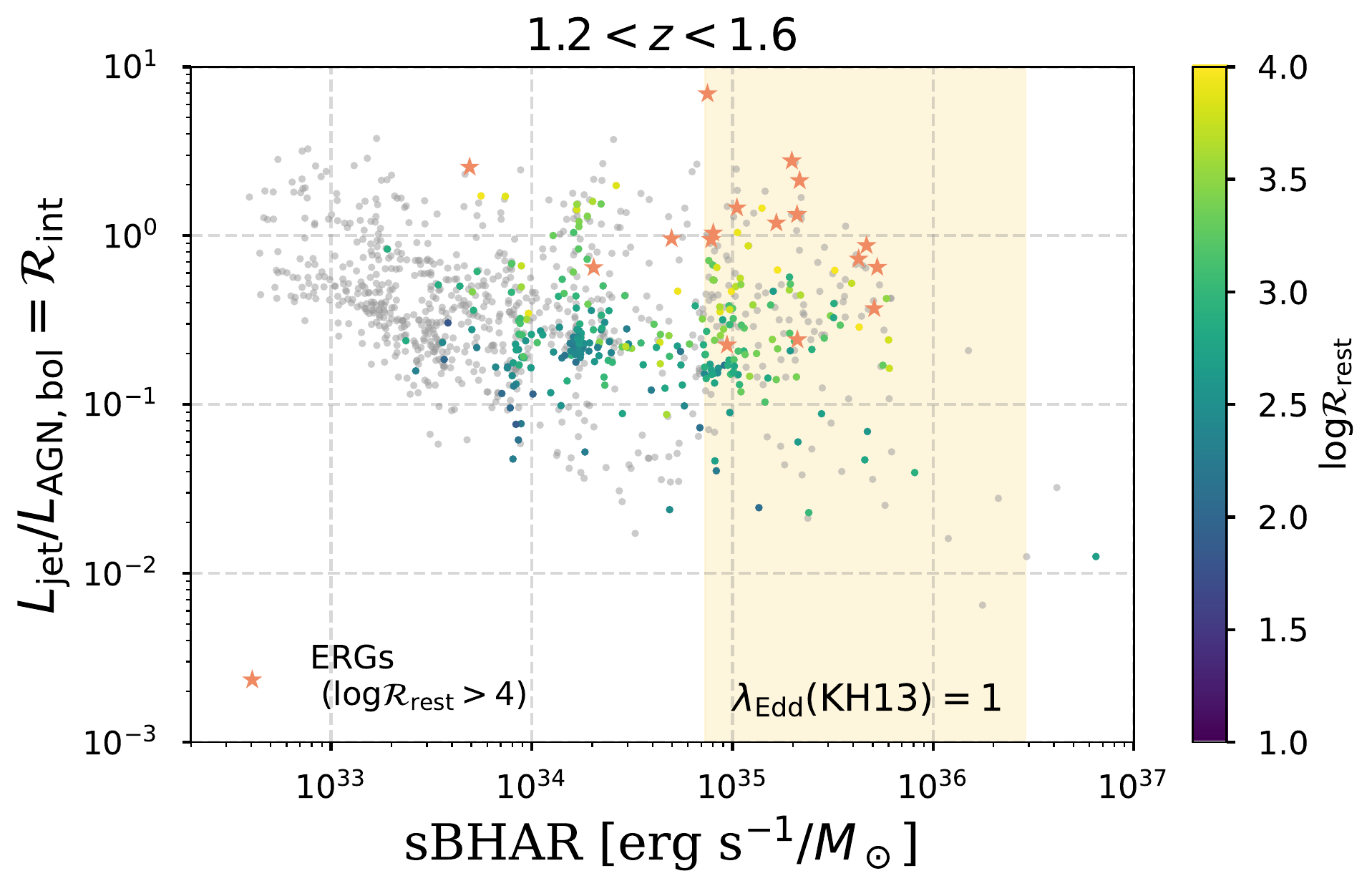}\\
\caption{
The same relations but with redshift-divided sample of $0.3<z<0.8$ (Left), $0.8<z<1.2$ (Center), and $1.2<z<1.6$ (Right) overlapped with the total sample ($0.3<z<1.6$) in gray.
(Top) The relation between SFR and
stellar-mass ($\mstar$) and the symbols are same with Figure~\ref{fig:SFRMstar}.
(Middle)
The relation between sBHAR and $\mstar$,
same with Figure~\ref{fig:sBHARvsMstar}.
(Bottom)
The relationship between $\ljet/\lbol = \Rint$ and sBHAR, same with Figure~\ref{fig:R_vs_sBHAR}.
}\label{fig:highz}
\end{center}
\end{figure*}

The physical properties between \ERGs\ and \NRGs\ are explored in Section~\ref{sec:analysis} and discussed in Section~\ref{sec:discussion}. As shown in Figure~\ref{fig:L_vs_z},
\NRGs\ distribute smoothly in a wide redshift range
at $0.3<z<1.6$, while \ERGs\ are slightly clustered at $z>0.8$. 
Thus, one might wonder whether the difference in physical parameters between \ERGs\ and \NRGs\ is caused by the redshift difference.
We here split the sample into the three redshift bins
($z=0.3$--$0.8$, $0.8$--$1.2$, and $1.2$--$1.6$)
for mitigating such redshift dependence and then compare the key parameters of the two populations.

The top panels show the relations between SFR and $\mstar$, which are similar to ones with Figure~\ref{fig:SFRMstar}.
Most of the \NRGs\ reside below the MS in the $0.3<z<0.8$, then the fraction of sources on and above the MS increases as a function of redshift. On the other hand, \ERGs\ always locate the lower $\mstar$ range in each redshift bin and most of them are above the MS (except several sources on and below the MS).
Considering that the stellar-mass is estimated from the photometries mainly in the optical or near-IR bands, the deeper Subaru/HSC photometreis enable us to find this smaller stellar-mass population as shown in Figure~\ref{fig:Mstar_vs_L1p4}.

The middle panels show the relations between sBHAR and $\mstar$, which are the redshift-divided version of Figure~\ref{fig:sBHARvsMstar}.
At $0.3<z<0.8$, most of the \NRGs\ are located at low sBHAR and high $\mstar$ cluster region where 
$\mathrm{sBHAR}/\sBHARunit \sim10^{33.5}$ and $\mstar \sim 10^{11.5}\msun$, and then they move to higher sBHAR sequence with redshift.
On the other hand, \ERGs\ always locate at the top edge of the sequence,
keeping high sBHAR values and sometimes reaching the sBHAR equivalent to $\lambdaedd\sim1$.

Finally, the bottom panels show the relation between $\Rint$ and sBHAR, which are the redshift-divided ones of Figure~\ref{fig:R_vs_sBHAR}.
Most of the \NRGs\ are located at the top left region at $0.3<z<0.8$, and then the population moves to higher sBHAR and smaller $\Rint$ sequence, moving to bottom-center or bottom-right of the panel. 
This is a consistent view of local radio galaxies at lower-$z$ and radio quasar population at higher-$z$ as discussed in Section~\ref{sec:figure7}.
On the other hand, \ERGs\ reside in the top right part of the plane,
and does not show a redshift evolution, because of its
extreme selection cut of $\log \R > 4$, limiting the sample only in the top right area of the panel.




\bibliographystyle{aasjournal}
\bibliography{ms}



\end{document}